\documentstyle{report}

\input{epsf}

\begin{document}

\thispagestyle{empty}

\begin{flushright}
{\small LAPTH-Conf-1067/04}
\end{flushright}

\vspace{3cm}

\begin{center}

{\Large \bf
An overview of Cosmology}\footnote{
These notes were prepared for the 2002, 2003 and 2004 sessions of the
Summer Students Programme of CERN. Most of them was written when I
was a Fellow
in the Theoretical Physics Division, CERN, CH-1211 Geneva 23 (Switzerland).
}

\vspace{0.3cm}

Julien Lesgourgues
\\
{\small \it 
LAPTH, Chemin de Bellevue, B.P. 110, F-74941 Annecy-Le-Vieux Cedex, France
}
\\
{\small (September 17, 2004)}
\end{center}

\vspace{2cm}

What is the difference between astrophysics and cosmology?  
While astrophysicists study the surrounding celestial bodies, like planets,
stars, galaxies, clusters of galaxies, gas clouds, etc., cosmologists
try to describe the evolution of the Universe as a whole, on the
largest possible distances and time scales. While purely philosophical
in the early times, and still very speculative at the beginning of
the twentieth century, cosmology has gradually entered into the realm of
experimental science over the past eighty years. Today, as we will see in
chapter two, 
astronomers are even able to obtain very precise maps of the
surrounding Universe a few billion years ago.

\vspace{0.2cm}

Cosmology has raised some fascinating questions like: is the Universe static
or expanding ? How old is it and what will be its future evolution ? 
Is it flat, open or closed ? Of what type of matter is it  composed ?
How did structures like galaxies form ? In this course, we will try
to give an overview of these questions, and of the partial answers
that can be given today.

\vspace{0.2cm}

In the first chapter, we will introduce some fundamental concepts, in
particular from General Relativity. Along this chapter, we will remain in
the domain of abstraction and geometry. In the second chapter, we will
apply these concepts to the real Universe and deal with concrete
results, observations, and testable predictions.

\newpage

\thispagestyle{empty}

\mbox{}

\newpage

\tableofcontents

\chapter{The Expanding Universe}
\label{chap1}

\section{The Hubble Law}
\label{sec11}

\subsection{The Doppler effect}
\label{sec111}

At the beginning of the XX-th century, the understanding of the global 
structure of the Universe beyond the scale of the solar system was still 
relying on pure speculation. In 1750, with a remarkable intuition, 
Thomas Wright noticed 
that the luminous stripe observed in the night sky and called the Milky 
Way could be a consequence of the spatial distribution of stars: they 
could form a thin plate, what we call now a galaxy. At that time, 
with the help of telescopes, many faint and diffuse objects had been 
already observed and listed, under the generic name of nebulae - in 
addition to the Andromeda nebula which is visible by eye, and has been known 
many centuries before the invention of telescopes. Soon after the 
proposal of Wright, the philosopher Emmanuel Kant suggested that some of 
these nebulae could be some other clusters of stars, far outside
the Milky Way. So, the idea of a galactic structure appeared in the mind 
of astronomers during the XVIII-th century, but even in the following century 
there was no way to check it on an experimental basis.

At the beginning of the nineteenth century, some physicists observed 
the first spectral lines. In 1842, Johann Christian Doppler argued that
if an observer receives a wave emitted by a body in motion, 
the wavelength that he will measure 
will be shifted proportionally to the speed of the emitting body 
with respect to the observer (projected along the line of sight):
\begin{equation}
\Delta \lambda / \lambda = \vec{v}.\vec{n} / c
\end{equation}
where $c$ is the celerity of the wave (See figure 1.1).
He suggested that this effect could
be observable for sound waves, and maybe also for light. The later assumption
was checked experimentally in 1868 by Sir William Huggins, who
found that the spectral lines of some neighboring stars were 
slightly shifted toward the red or blue ends of the spectrum.
So, it was possible to know the
projection along the line of sight of star velocities, $v_r$, using
\begin{equation}
z \equiv \Delta \lambda / \lambda = v_r / c
\end{equation}
where $z$ is called the redshift (it is negative in case of blue-shift) and
$c$ is the speed of light. Note that the redshift gives no indication 
concerning the distance of the star. At the beginning of the XX-th century,
with increasingly good instruments, people could also measure the redshift
of some nebulae. The first measurements, performed on the brightest objects, 
indicated 
some arbitrary distribution of red and blue-shifts, like for stars.
Then, with more observations, it appeared that the statistics was biased in 
favor of red-shifts, suggesting that a majority of nebulae were going 
away from us, unlike stars. This was raising new questions concerning
the distance and the nature of nebulae.
\begin{figure}[!bt]
\begin{center}
\epsfysize=3cm
\epsfbox{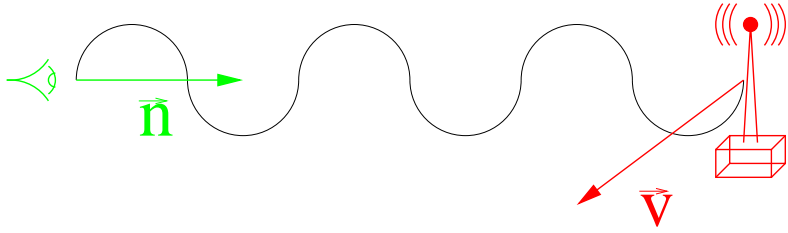}
\end{center}
\caption{The Doppler effect}
\end{figure}
\label{fig1}

\subsection{The discovery of the galactic structure}
\label{sec112}

In the 1920's, Leavitt and Shapley studied some particular stars, 
called the cepheids, known to have a periodic
time-varying luminosity. They could show that the period of cepheids
is proportional to their absolute luminosity $L$ (the absolute luminosity is
the total amount of light emitted by unit of time, i.e., the flux integrated 
on a closed surface around the star). They were also able to give the 
coefficient of proportionality.
So, by measuring the apparent luminosity, i.e. the flux $l$
per unit of surface through an instrument pointing to the star,
it was easy to get the distance of the star $r$ from 
\begin{equation}
L = l \times \left( 4 \pi r^2 \right)~.
\end{equation}
Using this technique, it became possible to measure the distance of 
various cepheids inside our galaxies, and to obtain the first 
estimate of the characteristic size of the stellar disk of the Milky Way
(known today to be around 80.000 light-years).

But what about nebulae? In 1923, the 2.50m telescope of Mount 
Wilson (Los Angeles) allowed Edwin Hubble to make the first observation 
of individual stars inside the brightest nebula, Andromeda. Some of these
were found to behave like cepheids, leading Hubble to give an
estimate of the distance of Andromeda. He found approximately 900.000 
light-years (but later, when cepheids were known better,
this distance was established to be around 2 million 
light-years). That was the first confirmation of the galactic structure
of the Universe: some nebulae were likely to be some distant replicas of 
the Milky Way, and the galaxies were separated by large voids.

\subsection{The Cosmological Principle}
\label{sec113}

This observation, together with the fact that most nebulae are
redshifted (excepted for some of the nearest ones like Andromeda), was
an indication that on the largest observable scales, the Universe was 
expanding. At the beginning, this idea was not widely accepted. Indeed, 
in the most general case, a given dynamics of expansion takes place around 
a center. Seeing
the Universe in expansion around us seemed to be an evidence for the
existence of a center in the Universe, very close to our own galaxy.

Until the middle age, the Cosmos was thought to be organized around
mankind, but the common wisdom of modern science suggests that there should 
be nothing special about the region or the galaxy in which we leave. This
intuitive idea was formulated by the astrophysicist Edward Arthur Milne
as the ``Cosmological Principle'': the Universe as a whole should be
homogeneous, with no privileged point playing a particular role.

Was the apparently observed expansion of the Universe a proof against
the Cosmological Principle? Not necessarily. The homogeneity
of the Universe is compatible either with a static distribution of galaxies,
or with a very special velocity field, obeying to a linear distribution:
\begin{equation}
\vec{v} = H~ \vec{r} 
\end{equation}
where $\vec{v}$ denotes the velocity of an arbitrary body with
position $\vec{r}$, and $H$ is a constant of proportionality.
An expansion described by this law is still homogeneous because it is left
unchanged by a change of origin. 
To see this, one can make an analogy with an infinitely large rubber grid, 
that would be stretched equally in all directions: it would expand, but with 
no center (see figure 1.2).
\begin{figure}[!bt]
\begin{center} 
\epsfysize=4cm
\epsfbox{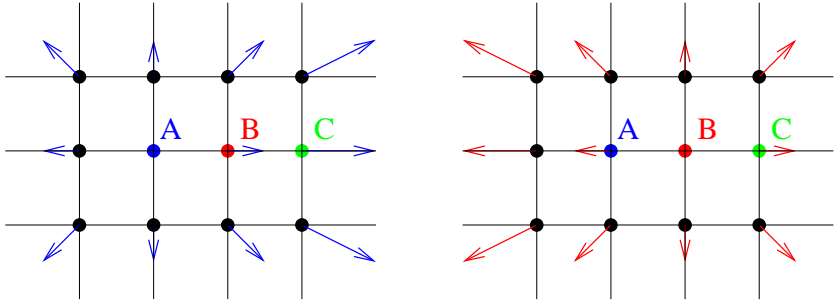}
\end{center}
\caption{
Homogeneous expansion on a two-dimensional grid.
Some equally-spaced observers are located at each intersection. 
The grid is plotted twice. On the left, the arrays show the expansion 
flow measured by A; on the right, the expansion flow measured
by B.
If we assume that the expansion is homogeneous, we get that
A sees B going away at the same velocity as B sees C going away.
So, using the additivity of speeds, the velocity of C with respect to
A must be twice the velocity of B with respect to A.
This shows that there is a linear relation between speed and distance,
valid for any observer. 
} 
\end{figure} 
\label{fig2}
This result is not true for any other velocity field.
For instance, the expansion law
\begin{equation}
\vec{v} = H~ |\vec{r}|~ \vec{r} 
\end{equation}
is not invariant under a change of origin: so, it has a center.

\subsection{Hubble's discovery}
\label{sec114}

So, a condition for the Universe to respect the Cosmological 
Principle is that the speed of galaxies along the line of sight, or
equivalently, their redshift, should be proportional to their distance.
Hubble tried to check this idea, still using the cepheid technique. He 
published in 1929 a study based on 18 galaxies, for which he had
measured both the redshift and the distance. His results were
showing roughly a linear relation between redshift and distance
(see figure 1.3).
He concluded that the Universe was in homogeneous expansion, and gave
the first estimate of the coefficient of proportionality $H$, called the 
Hubble parameter.

This conclusion has been checked several time with increasing precision and 
is widely accepted today. It can be considered as the starting point of
experimental cosmology. It is amazing to note that the data used by Hubble was
so imprecise that Hubble's conclusion was probably a bit biaised... 
\begin{figure}[!bt]
\begin{center}
\epsfysize=15cm
\epsfbox{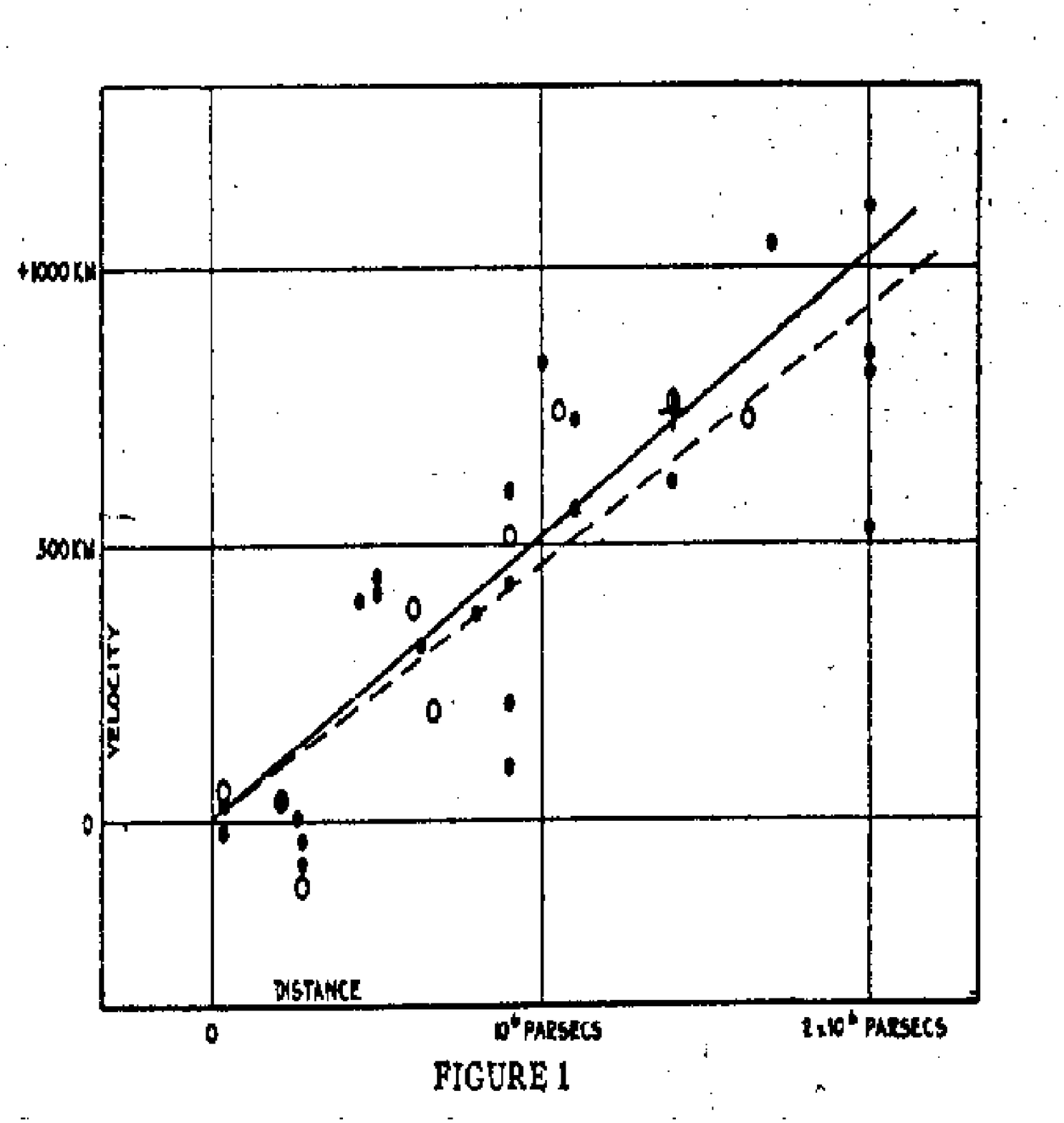}
\end{center}
\vspace{-3cm}
\label{fig3}
\caption{The diagram published by Hubble in 1929. The labels of the
horizontal (resp. vertical) axis are 0, 1, 2~Mpc (resp. 0,
500, 1000~km.s$^{-1}$). Hubble estimated the expansion rate to be
500~km.s$^{-1}$Mpc$^{-1}$. Today, it is known to be around
70~km.s$^{-1}$Mpc$^{-1}$.
}
\end{figure}
Anyway, current data leaves no doubt about the proportionality, even if
there is still an uncertainty concerning the exact value
of $H$. The Hubble constant is generally parametrized as 
\begin{equation}
H = 100~ h ~{\rm km~s}^{-1} {\rm Mpc}^{-1}
\end{equation}
where $h$ is the dimensionless ``reduced Hubble parameter'', 
currently known to be in the range $h = 0.71 \pm 0.04$, and Mpc denotes
a Mega-parsec, the unity of distance usually employed for cosmology
(1 Mpc $\simeq 3 \times 10^{22} {\rm m} \simeq 3 \times 10^{6}$ light-years;
the proper definition of a parsec is ``the distance to an object with a 
parallax of one arcsecond''; the parallax being half the angle under which a 
star appears to move when the earth makes one rotation around the sun).
So, for instance, a galaxy located at 10 Mpc goes away at a speed
close to 700 km~s$^{-1}$.

\subsection{Homogeneity and inhomogeneities}
\label{sec115}

Before leaving this section, we should clarify one point about the
``Cosmological Principle'', i.e., the assumption that the Universe
is homogeneous. Of course, nobody has ever claimed that the Universe was 
homogeneous on small scales, since compact objects like planets or
stars, or clusters of stars like galaxies are inhomogeneities in themselves. 
The Cosmological
Principle only assumes homogeneity after smoothing over some characteristic 
scale. By analogy, take a grid of step $l$
(see figure 1.4), and put one object in each
intersection, with a randomly distributed mass (with all masses obeying 
to the same distribution of probability). Then, make a random
displacement of each object (again with all displacements obeying to the 
same distribution of probability). At small scales, the mass density is 
obviously inhomogeneous for three reasons: the objects are compact, they
have different masses, and they are separated by different distances.
However, since
the distribution has been obtained by performing a random shift 
in mass and position,
starting from an homogeneous structure, it is clear even intuitively that 
the mass density smoothed over some large scale will remain 
homogeneous again.
\begin{figure}[!bt]
\begin{center}
\epsfysize=6cm
\epsfbox{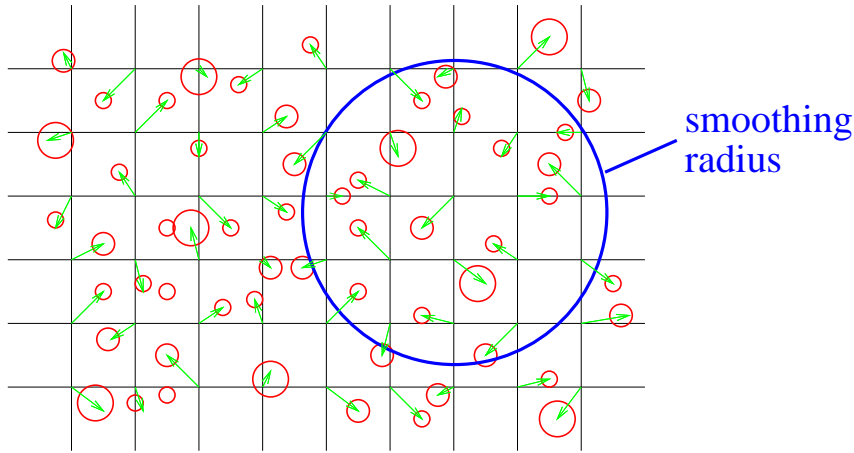}
\end{center}
\label{fig4}
\caption{
We build an inhomogeneous distribution of objects in the following
way: starting from each intersection of the grid, we draw a random
vector and put an object of random mass at the extremity of the vector.
Provided that all random vectors and masses obey to the same distributions
of probability, the mass density is still homogeneous when it is smoothed over
a large enough smoothing radius (in our example, the typical
length of the vectors 
is smaller than the step of the grid; but our conclusion would 
still apply if the vectors were larger than the grid step,
provided that the smoothing radius is even larger). 
This illustrates the concept of homogeneity above a given scale, like 
in the Universe.
}
\end{figure}

The Cosmological Principle should be understood in this sense.
Let us suppose that the Universe is almost homogeneous at a scale 
corresponding, say, to the typical intergalactic
distance, multiplied by thirty or so. 
Then, the Hubble law doesn't have to be verified exactly for an individual 
galaxy, because of peculiar motions resulting from the fact that galaxies 
have slightly different masses, and are not in a perfectly ordered phase like 
a grid. But the Hubble law should be verified in average, provided that the
maximum scale of the data is not smaller than the scale of homogeneity. The 
scattering of the data at a given scale reflects the level of inhomogeneity, 
and when using data on larger and larger scales, the scattering must be less 
and less significant. This is exactly what is observed in practice. 
An even better proof of the homogeneity of
the Universe on large scales comes from the Cosmic Microwave Background,
as we shall see in section \ref{sec22}.

We will come back to these issues in section \ref{sec22}, and show how
the formation of inhomogeneities on small scales are
currently understood and quantified within some precise physical
models.

\section{The Universe Expansion from Newtonian Gravity}
\label{sec12}

It is not enough to observe the galactic motions, one should 
also try to explain it with the laws of physics.

\subsection{Newtonian Gravity versus General Relativity}
\label{sec121}

On cosmic scales, the only force expected to be relevant is gravity.
The first theory of gravitation, derived by Newton, was embedded later
by Einstein into a more general theory: General Relativity 
(thereafter denoted GR).
However, in simple words, GR is relevant only for describing gravitational
forces between bodies which have relative motions comparable to the speed of 
light\footnote{Going a little bit more into details, it is also relevant
when an object is so heavy and so close that the speed of liberation from
this object 
is comparable to the speed of light.}.
In most other cases, Newton's gravity gives a sufficiently accurate
description.

The speed of neighboring galaxies 
is always much smaller than the speed of light.
So, {\it a priori}, 
Newtonian gravity should be able to explain the Hubble flow.
One could even think that historically, Newton's law led to 
the prediction of the Universe expansion, or
at least, to its first interpretation.
Amazingly, and for reasons which are more mathematical than physical,
it happened not to be the case: the first attempts to describe
the global dynamics of the Universe came with GR, in the 1910's.
In this course, for pedagogical purposes, we will not follow the
historical order, and start with the Newtonian approach.

Newton himself did the first step in the argumentation. He noticed that
if the Universe was of finite size, and governed by the law of gravity,
then all massive bodies would unavoidably
concentrate into a single point, just because of 
gravitational attraction. If instead it was infinite, 
and with an approximately homogeneous distribution at initial time, 
it could concentrate
into several points, like planets and stars,
because there would be no center to fall in. 
In that case, the motion of each massive body would be driven by 
the sum of an infinite number of gravitational forces.
Since the mathematics of that time didn't allow to deal with this situation,
Newton didn't proceed with his argument.

\subsection{The rate of expansion from Gauss theorem}
\label{sec122}

In fact, using Gauss theorem, this problem turns out to be quite
simple.  Suppose that the Universe consists in many massive bodies
distributed in an isotropic and homogeneous way (i.e., for any
observer, the distribution looks the same in all directions).  This
should be a good modelization of the Universe on sufficiently large
scales. We wish to compute the motion of a particle located at a
distance $r(t)$ away from us.  Because the Universe is assumed to be
isotropic, the problem is spherically symmetric, and we can employ
Gauss theorem on the sphere centered on us and attached to the
particule (see figure 1.5).
\begin{figure}[!bt]
\begin{center}
\epsfysize=5cm
\epsfbox{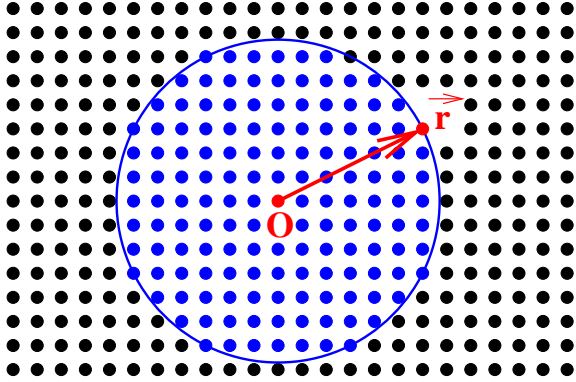}
\end{center}
\label{fig5}
\caption{Gauss theorem applied to the local Universe.}
\end{figure}
The acceleration of any particle on the surface of this sphere reads
\begin{equation}
\ddot{r}(t) = - \frac{{\cal G} M(r(t))}{r^2(t)}
\label{newton}
\end{equation}
where ${\cal G}$ is Newton's constant and
$M(r(t))$ is the mass contained inside the sphere of radius $r(t)$. 
In other words, the particle feels the same force as if it had a
two-body interaction with the mass of the sphere concentrated at the 
center.
Note that $r(t)$ varies with time, but $M(r(t))$ remains constant: because of 
spherical symmetry, no particle can enter or leave the sphere, which 
contains always the same mass.

Since Gauss theorem allows us to make completely abstraction of the mass
outside the sphere\footnote{The argumentation that we present 
here is useful for guiding
our intuition, but we should say that it is not fully self-consistent.
Usually, when we have to deal with a spherically symmetric mass distribution,
we apply Gauss theorem inside a sphere, and forget completely
about the external mass. This is actually not correct when
the mass distribution spreads out to infinity. Indeed, in our example,
Newtonian gravity implies that
a point inside the sphere would
feel all the forces from all bodies inside and outside the sphere, 
which would exactly 
cancel out. Nevertheless, the present calculation based 
on Gauss theorem does lead to a correct prediction for the expansion
of the Universe. In fact, this can be rigorously justified only
{\it a posteriori}, after a full general relativistic study. In GR, 
Gauss theorem can be generalized thanks to Birkhoff's
theorem, which is valid also when the mass distribution spreads to infinity.
In particular, for an infinite spherically symmetric
matter distribution, Birkhoff's theorem says that we can isolate
a sphere as if there was nothing outside of it. Once this formal step
has been performed, nothing prevents us from using Newtonian gravity
and Gauss theorem inside a smaller sphere, as if the external
matter distribution was 
finite. This argument justifies rigorously the calculation of this section.}, 
we can make an analogy with the motion e.g. of a 
satellite ejected vertically from the Earth. We know
that this motion depends on the initial velocity, compared with the
speed of liberation from the Earth: if the initial speed is large
enough, the satellites goes away indefinitely, 
otherwise it stops and falls down. We can see this 
mathematically by multiplying equation (\ref{newton}) 
by $\dot{r}$, and integrating it over time:
\begin{equation}
\frac{\dot{r}^2(t)}{2} = \frac{{\cal G} M(r(t))}{r(t)} - \frac{k}{2}
\end{equation}
where $k$ is a constant of integration. We can replace the mass $M(r(t))$
by the volume of the sphere multiplied by the homogeneous
mass density $\rho_{\rm mass}(t)$, and rearrange the equation as
\begin{equation}
\left( \frac{\dot{r}(t)}{r(t)} \right)^2
= \frac{8 \pi {\cal G}}{3} \rho_{\rm mass}(t) - \frac{k}{r^2(t)}~.
\label{newton.expansion}
\end{equation}
The quantity $\dot{r} / r$ is called the rate of expansion.
Since $M(r(t))$ is time-independent, the mass density 
evolves as $\rho_{\rm mass}(t) \propto r^{-3}(t)$ (i.e., matter is simply
diluted when the Universe expands).
The behavior of $r(t)$ depends on the sign of $k$. If $k$ is positive,
$r(t)$ can grow at early times but it always decreases at late times, 
like the altitude of the
satellite falling back on Earth: this would correspond to
a Universe expanding first, and then collapsing. 
If $k$ is zero or negative, the expansion lasts forever.
\begin{figure}[!bt]
\begin{center}
\epsfysize=6cm
\epsfbox{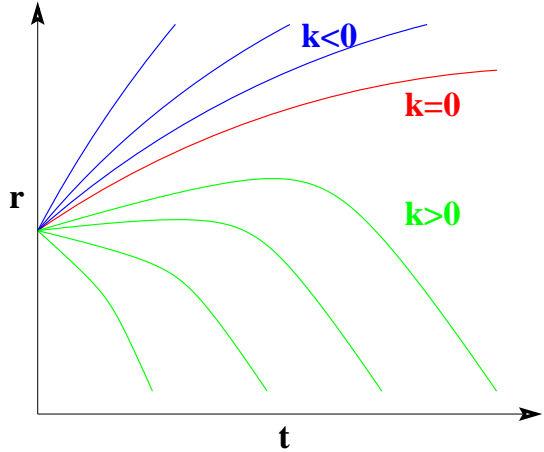}
\end{center}
\label{fig6}
\caption{The motion of expansion in a Newtonian Universe
is equivalent to that of a body ejected from Earth. It depends
on the initial rate of expansion compared with a critical density.
When the parameter $k$ is zero or
negative, the expansion lasts forever, otherwise the Universe 
re-collapses ($r \rightarrow 0$).
}
\end{figure}

In the case of the satellite, the critical value, which is the
speed of liberation (at a given altitude), depends on the mass of the Earth.
By analogy, in the case of the Universe, the important quantity
that should be compared with some critical value
is the homogeneous mass density. If at all times
$\rho_{\rm mass}(t)$ is bigger than the critical value
\begin{equation}
\rho_{\rm mass}(t) = \frac{3 (\dot{r}(t)/r(t))^2 }{8 \pi {\cal G}} 
\end{equation}
then $k$ is positive and the Universe will re-collapse.
Physically, it means that
the gravitational force wins against inertial effects.
In the other case, the Universe expands forever, because the density is
too small with respect to the expansion velocity, and gravitation
never takes over inertia. The case $k=0$ corresponds to a kind
of equilibrium between gravitation and inertia in which the Universe
expands forever, following a power--law: $r(t) \propto t^{2/3}$.

\subsection{The limitations of Newtonian predictions}
\label{sec123}

In the previous calculation, we cheated a little bit: we assumed that
the Universe was isotropic around us, but we didn't check that it was
isotropic everywhere (and therefore homogeneous). Following what we
said before, homogeneous expansion requires proportionality between
speed and distance at a given time.  Looking at equation
(\ref{newton.expansion}), we see immediately that this is true only
when $k=0$. So, it seems that the other solutions are not compatible
with the Cosmological Principle. We can also say that if the Universe
was fully understandable in terms of Newtonian mechanics, then the
observation of linear expansion would imply that $k$ equals zero
and that there is a precise relation between the density and the
expansion rate at any time.

This argument shouldn't be taken seriously, because the link that
we made between homogeneity and linear expansion was based
on the additivity of speed (look for instance at the caption of figure
1.2), and therefore, on Newtonian mechanics. 
But Newtonian mechanics cannot be applied at large distances, where
$v$ becomes large and comparable to the speed of light.
This occurs around a characteristic scale
called the Hubble radius $R_H$ :
\begin{equation}
R_H = c H^{-1},
\end{equation}
at which the Newtonian expansion law gives $v = H R_H = c$.

So, the full problem has to be formulated in relativistic terms. In the
GR results, we will see again some solutions with $k \neq 0$, but
they will remain compatible with the homogeneity of the Universe.

\section{General relativity and the Friemann-Lema\^{\i}tre model}
\label{sec13}

It is far beyond the scope of this course to introduce General Relativity,
and to derive step by step the relativistic laws governing the evolution
Universe. We will simply write these laws, asking the reader
to admit them - and in order to give a perfume of the underlying 
physical concepts, we will comment on 
the differences with their Newtonian counterparts.

\subsection{The curvature of space-time}
\label{sec131}

When Einstein tried to build a theory of gravitation
compatible with the invariance of the speed of light, he found that 
the minimal price to pay was : 
\begin{itemize}
\item
to abandon the idea of a gravitational potential,
related to the distribution of matter, 
and whose gradient gives the gravitational field in any point.
\item 
to assume that our four-dimensional space-time is curved
by the presence of matter.
\item 
to impose that free-falling objects
describe geodesics in this space-time.
\end{itemize}
What does that mean in simple words?

First, let's recall briefly what a curved space is, first
with only two-dimensional surfaces. Consider a plane, a sphere and an 
hyperboloid. For us, it's obvious that the sphere and the hyperboloid are 
curved, because we can visualize them in our three-dimensional space:
so, we have an intuitive notion of what is flat and what is curved. But if
there were some two-dimensional people living on these surfaces,
not being aware of the existence of a third dimension, how
could they know whether they leave in a flat or a in curved space-time?

There are several ways in which they could measure it. One would be to obey 
the following prescription: walk in straight line on a distance $d$; turn 90 
degrees left; repeat this sequence three times again; see whether you are back 
at your initial position. 
The people on the three surfaces would find that they are back there
as long as they walk along a small square,
smaller than the radius of curvature.
But a good test is to repeat the operation on larger and 
larger distances.
When the size of the square will be of the same order of magnitude as the 
radius of curvature, the habitant of the sphere will notice that
before stopping, he crosses the first branch of his trajectory
(see figure 1.7).
The one on the hyperboloid will stop without closing his trajectory.
\begin{figure}[!bt]
\begin{center}
\epsfysize=3cm
\epsfbox{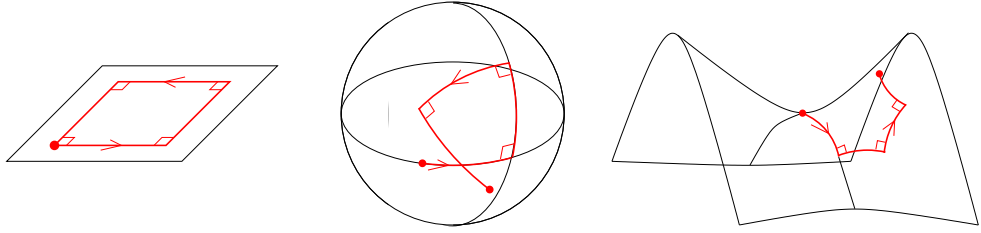}
\end{center}
\label{fig7}
\caption{Measuring the curvature of some two-dimensional spaces.  By
walking four times in straight line along a distance $d$, and turning
90 degrees left between each walk, a small man on the plane would find
that he is back at his initial point. Doing the same thing, a man on
the sphere would walk across his own trajectory and stop away from his
departure point. Instead, a man on the hyperboloid would not close his
trajectory.}
\end{figure}

It is easy to think of the curvature of a two-dimensional surface
because we can visualize it embedded into three-dimensional space. 
Getting an
intuitive representation of a three-dimensional curved space is much more 
difficult. A 3-sphere and a 3-hyperboloid could be defined analytically as
some 3-dimensional spaces obeying to the equation
$a^2 + b^2 + c^2 \pm d^2 = R^2$
inside a 4-dimensional Euclidian space with coordinates $(a,b,c,d)$.
If we wanted to define them by making use of only three dimensions,
the problem would be exactly like for drawing a planisphere of the Earth.
We would need to give a map of the space, together with a crucial
information: the scale of the map as a function of the location on the map
- the scale on a planisphere is not uniform!
This would bring us to a mathematical formalism called Riemann geometry,
that we don't have time to introduce here.

That was still for three dimensions. 
The curvature of a four-dimensional space-time is impossible to
visualize intuitively, first because it has even more dimensions, 
and second because even in special/general relativity, there is
a difference between time and space
(for the readers who are familiar with special relativity, 
what is referred here is the negative signature of the metric).

The Einstein theory of gravitation says that four-dimensional space-time is 
curved, and that the curvature in each point is given entirely in terms 
of the matter content in this point. In simple words, 
this means that the curvature plays more or less the same role as the 
potential in Newtonian gravity. But the potential was simply
a function of space and time coordinates. In GR, the full curvature 
is described not by a function, but by something more complicated
- like a matrix of functions obeying to particular laws - called a tensor.

Finally, the definition of geodesics
(the trajectories of free-falling bodies) is the following. 
Take an initial point and an initial direction. They define a 
unique line, 
called a geodesic, such that any segment of the line 
gives the shortest trajectory between the two points
(so, for instance, on a sphere of radius $R$, the geodesics are all 
the great circles of radius $R$, and nothing else).
Of course, geodesics depend on curvature. All free-falling bodies
follow geodesics, including light rays. This leads for instance to
the phenomenon of gravitational lensing (see figure 1.8).
\begin{figure}[!bt]
\begin{center}
\epsfysize=5cm
\epsfbox{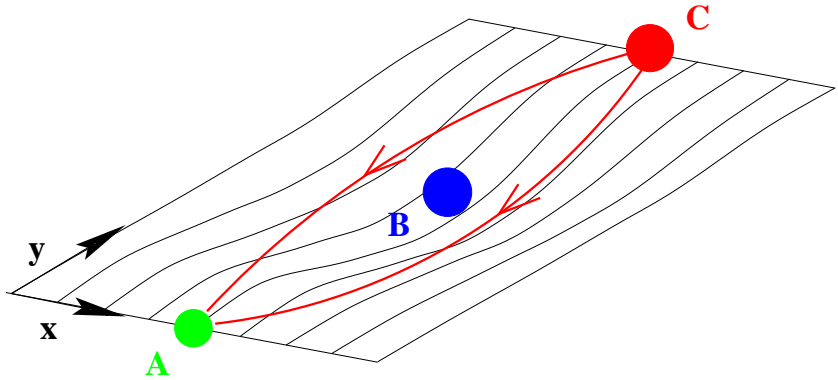}
\end{center}
\label{fig8}
\caption{Gravitational lensing. Somewhere between an object $C$
and an observer $A$, a massive object $B$
- for instance, a galaxy - curves its surrounding space-time.
Here, for simplicity, we only draw two spatial dimensions.
In absence of gravity and curvature, the only possible trajectory of light
between $C$ and $A$ would be a straight line. But because of
curvature, the straight line is not anymore the shortest trajectory.
Photons prefer to follow two geodesics, symmetrical around $B$.
So, the observer will not see one image of $C$, but two distinct
images. In fact, if we restore the third spatial
dimension, and if the three points are perfectly aligned, the image of $C$
will appear as a ring around $B$. This phenomenon is observed in practice.
}
\end{figure}

So, in General Relativity, gravitation is not formulated as a force
or a field, but as a curvature of space-time, sourced by matter. 
All isolated systems follow geodesics which are bent
by the curvature. In this way, their trajectories are affected by
the distribution of matter around them: this is precisely what
gravity means.

\subsection{Building the first cosmological models}
\label{sec132}

After obtaining the mathematical formulation of General Relativity,
around 1916, Einstein studied various testable consequences of his
theory in the solar system (e.g., corrections to the trajectory of
Mercury, or to the apparent diameter of the sun during an eclipse).
But remarkably, he immediately understood that GR could also be
applied to the Universe as a whole, and published some first
attempts in 1917.  However, Hubble's results concerning the
expansion were not known at that time, and most physicists had the
prejudice that the Universe should be not only isotropic and
homogeneous, but also static -- or stationary. As a consequence, 
Einstein (and other people like De Sitter) found some interesting 
cosmological solutions, but not the ones that really describe our 
Universe.

A few years later, some other physicists tried to relax the assumption 
of stationarity. The first was the russo-american Friedmann (in 1922), 
followed closely by the Belgian physicist and priest Lema\^{\i}tre
(in 1927), and then by some americans, Robertson and Walker. 
When the Hubble flow was discovered in 1929, it became clear for a
fraction of the scientific community that the Universe could be described 
by the equations of Friedmann, Lema\^{\i}tre, Roberston and Walker.
However, many people -- including Hubble and Einstein themselves -- remained 
reluctant to this idea for many years. Today, the Friedmann -- Lema\^{\i}tre
model is considered as one of the major achievements of the XXth century.

Before giving these equations, we should stress that they look pretty
much the same as the Newtonian results given above - although some
terms seem to be identical, but have a different physical interpretation. 
This similarity was noticed only much later. Physically,
it has to do with the fact that there exists a 
generalization of Gauss theorem to GR, known as Birkhoff theorem.
So, as in Newtonian gravity, one can study the expansion 
of the homogeneous Universe by considering only matter inside
an sphere. But in small regions, General Relativity admits
Newtonian gravity as an asymptotic limit, and so 
the equations have many similarities.

\subsection{Our Universe is curved}
\label{sec133}

The Friemann-Lema\^{\i}tre model is defined as the most general 
solution of the laws of General Relativity,
assuming that the Universe is isotropic and homogeneous.
We have seen that in GR, matter curves space-time. So, the
Universe is curved by its own matter content, 
along its four space and time dimensions. However, because
we assume homogeneity, we can decompose the total curvature into
two  parts:
\begin{itemize}
\item the spatial curvature, i.e., the curvature of the usual
  3-dimensional space $(x,y,z)$ at fixed time. This curvature
  can be different at different times. It is maximally symmetric,
  i.e., it makes no difference between the three directions $(x,y,z)$.
  There are only three maximally symmetric solutions: space can be
  Euclidean, a 3-sphere with finite volume, or a 3-hyperboloid. 
  These three possibilities are referred to as a flat, a closed 
  or an open Universe. {\it A priori}, nothing
  forbids that we leave in a closed or in an open Universe: if the
  radius of curvature was big enough, say, bigger than the size of our
  local galaxy cluster, then the curvature would show up only
  in long range astronomical observations. In the next chapter, we will 
  see how recent observations are able to give a precise
  answer to this question.
\item the two-dimensional space-time curvature, i.e., for instance,
  the curvature of the $(t,x)$ space-time. Because of isotropy,
  the curvature of the $(t,y)$ and the $(t,z)$ space-time have to be the 
  same. This curvature is the one responsible for the expansion.
  Together with the spatial curvature, it fully describes gravity in
  the homogeneous Universe. 
\end{itemize}
The second part is a little bit more difficult to understand
for the reader who is not familiar with GR, and causes 
a big cultural change in one's intuitive understanding of space and time. 

\subsection{Comoving coordinates}
\label{sec134}

In the spirit of General Relativity, we will consider a set
of three variables that will represent the spatial coordinates (i.e.,
a way of mapping space, and of giving a label to each point), 
but NOT directly a measure of distances!

The laws of GR allow us to work with any system of coordinates that we
prefer. For simplicity, in the Friedmann model, people 
generally employ a particular system
of spherical coordinates $(r, \theta, \phi)$ called ``comoving coordinates'', 
with the following striking property:
the physical distance $dl$ between two infinitesimally
close objects with coordinates $(r, \theta, \phi)$ and 
$(r+dr, \theta + d \theta, \phi + d\phi)$ 
is not given by
\begin{equation}
dl^2 = dr^2 + r^2 (d \theta^2 + \sin^2 \! \theta ~d \phi^2)
\label{euclidian.distance}
\end{equation}
as in usual Euclidean space, but 
\begin{equation}
dl^2 = a^2 \! (t) 
\left[ \frac{dr^2}{1 - k r^2} 
+ r^2 (d \theta^2 + \sin^2 \! \theta ~d \phi^2 ) \right]
\label{friedman.distance}
\end{equation}
where $a(t)$ is a function of time, called the scale factor,
whose time-variations account for the curvature
of two-dimensional space-time; 
and $k$ is a constant number, related to the spatial
curvature: if $k=0$, the Universe is flat, if $k>0$, it is closed,
and if $k<0$, it is open. In the last two cases, the radius of curvature
$R_c$ is given by
\begin{equation}
R_{c}(t) = \frac{a(t)}{\sqrt{|k|}}.
\end{equation}
When the Universe is closed, it has a finite volume, so the coordinate
$r$ is defined only up to a finite value: $0 \leq r < 1 /\sqrt{k} $.

If $k$ was equal to zero and $a$ was constant in time, 
we could redefine the coordinate system with 
$(r', \theta', \phi') = (a r, \theta, \phi)$, find the usual expression
(\ref{euclidian.distance}) for distances, and go back to Newtonian mechanics. 
So, we stress again that
the curvature really manifests itself as $k \neq 0$ (for spatial curvature)
and $\dot{a} \neq 0$ (for the remaining space-time curvature). 

Note that when $k=0$, we can rewrite $dl^2$ in Cartesian coordinates:
\begin{equation}
dl^2 = a^2 \! (t) \left( dx^2 + dy^2 + dz^2 \right)~.
\end{equation}
In that case, the expressions for $dl$ contains only the differentials
$dx$, $dy$, $dz$, and is obviously left unchanged by a change of
origin - as it should be, because we assume a homogeneous Universe.
But when $k \neq 0$, there is the additional factor $1/(1 - k r^2)$,
where $r$ is the distance to the origin. So, naively, one could think
that we are breaking the homogeneity of the Universe, and privileging
a particular point. This would be true in Euclidean geometry, but in
curved geometry, it is only an artifact.  Indeed, choosing a system of
coordinates is equivalent to mapping a curved surface on a plane. By
analogy, if one draws a planisphere of half-of-the-Earth by projecting
onto a plane, one has to chose a particular point defining the axis of
the projection. Then, the {\it scale} of the map is not the same all
over the map and is symmetrical around this point.
\begin{figure}[!bt]
\begin{center}
\epsfysize=10cm
\epsfbox{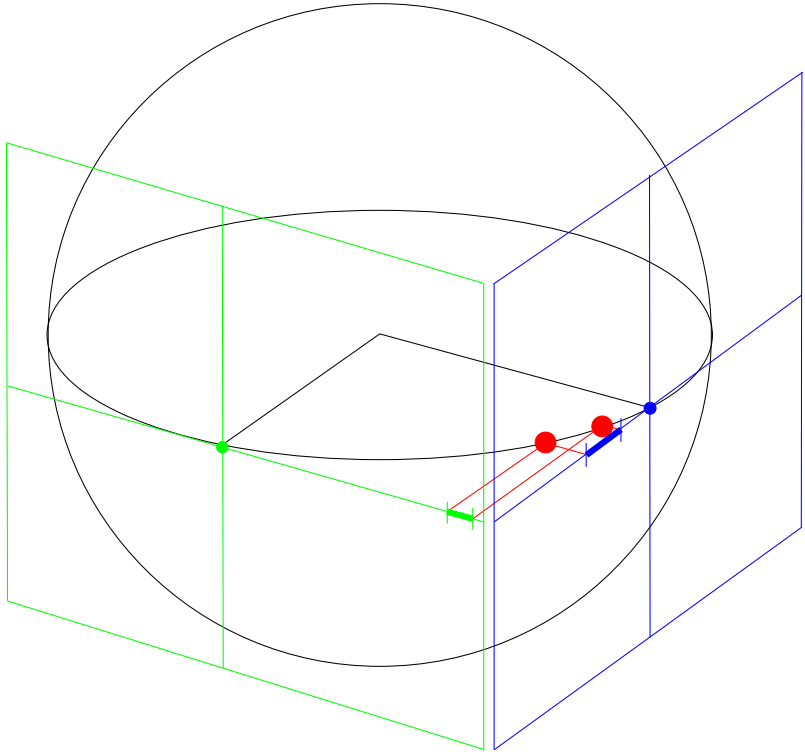}
\end{center}
\label{fig9}
\caption{Analogy between the Friedman coordinates for an open/closed
Universe, and the mapping of a sphere by parallel projection
onto a plane. Two maps are shown, with different axes of projection.
On each map, the scale depends on the distance to the center of the map: 
the scale is smaller at the center than on the borders. However,
the physical distance between too points on the sphere can be computed
equally well using the two maps and their respective scaling laws.}
\end{figure}
However, if one reads two maps with different projection axes (see
figure 1.9), and computes the distance between Geneva and Rome using
the two maps with their appropriate scaling laws, he will of course
find the same physical distance.  This is exactly what happens in our
case: in a particular coordinate system, the expression for physical
lengths depends on the coordinate {\it with respect to the origin},
but physical lengths are left unchanged by a change of coordinates.
In figure 1.10 we push the analogy and show how the factor $1/(1 - k
r^2)$ appears explicitly for the parallel projection of a 2-sphere
onto a plane.
\begin{figure}[!bt]
\begin{center}
\epsfysize=5cm
\epsfbox{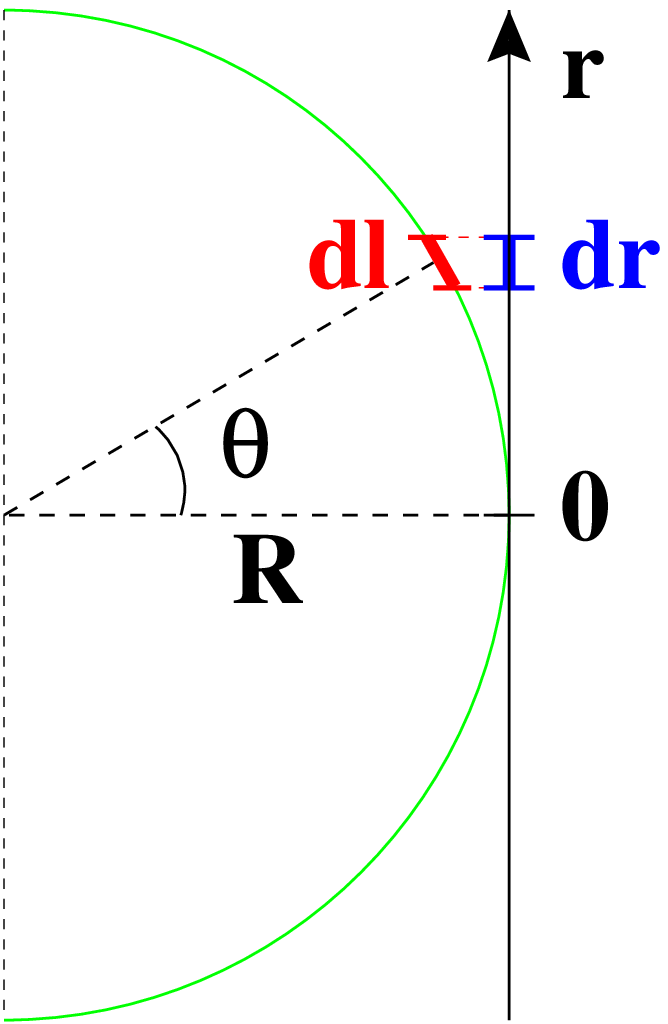}
\end{center}
\label{fig10}
\caption{
We push the analogy between the Friedmann coordinates in a closed
Universe, and the parallel projection of a half--sphere onto a plane.
The physical distance $dl$ on the surface of the sphere corresponds
to the coordinate interval $dr$, such that
$
dl 
= {dr} / {\cos \theta} 
= {dr} / {\sqrt{1 - \sin^2 \! \theta}}
= {dr} / {\sqrt{1 - (r/R)^2}}
$,
where $R$ is the radius of the sphere. This is exactly the same law
as in a closed Friedman Universe.
}
\end{figure}
The rigorous proof that equation (\ref{friedman.distance})
is invariant by a change of origin would be more complicated, but 
essentially similar.

Free-falling bodies follow geodesics in this curved space-time. Among
bodies, we can distinguish between photons, which are relativistic 
because they travel at the speed of light, and ordinary matter like 
galaxies, which are non-relativistic (i.e., an observer sitting on them 
can ignore relativity at short distance, as we do on Earth).

The geodesics followed by non-relativistic bodies in space-time
are given by $dl=0$, i.e., fixed spatial coordinates. So, 
galaxies are at rest in comoving
coordinates. But it doesn't mean that the Universe is static, because 
all distances grow proportionally to
$a(t)$: so, the scale factor accounts for the homogeneous expansion.
A simple analogy helps in understanding this subtle concept.
Let us take a rubber balloon and draw some points on the surface.
Then, we inflate the balloon. The distances between all the points grow
proportionally to the radius of the balloon. This is not because the points
have a proper motion on the surface, but because all the lengths
on the surface of the balloon increase with time.

The geodesics followed by photons are straight lines in 3-space, 
but not in space-time. Locally, they obey to the same
relation as in Newtonian mechanics: $c~ dt = dl$, i.e.,
\begin{equation}
c^2 dt^2 = 
a^2 \! (t) 
\left[ \frac{dr^2}{1 - k r^2} 
+ r^2 (d \theta^2 + \sin^2 \! \theta ~d \phi^2 ) \right]~.
\label{photon.propagation}
\end{equation}
So, on large scales,
the distance traveled by light in a given time interval is not
$\Delta l = c \Delta t$ like in Newtonian mechanics, but is given 
by integrating the infinitesimal equation of motion.
We can write this integral, taking
for simplicity a photon with constant $(\theta, \phi)$ coordinates
(i.e., the origin is chosen on the trajectory of the photon).
Between $t_1$ and $t_2$, the change in {\it comoving coordinates}
for such a photon is given by
\begin{equation}
\int_{r_1}^{r_2} \frac{dr}{\sqrt{1-kr^2}} 
= \int_{t_1}^{t_2} \frac{c}{a(t)} dt.
\end{equation}

This equation for the propagation of light is extremely important -
probably, one of the two most important of cosmology, together with
the Friedmann equation, that we will give soon. It is on the basis of
this equation that we are able today to measure the curvature of the
Universe, its age, its acceleration, and other fundamental quantities.

\subsection{Bending of light in the expanding Universe}
\label{sec135}

Lets give a few examples of the implications of 
equation (\ref{photon.propagation}),
which gives the bending of the trajectories followed by photons
in our curved space-time, as illustrated in figure 1.11.
\begin{figure}[!bt]
\begin{center}
\epsfysize=10cm
\epsfbox{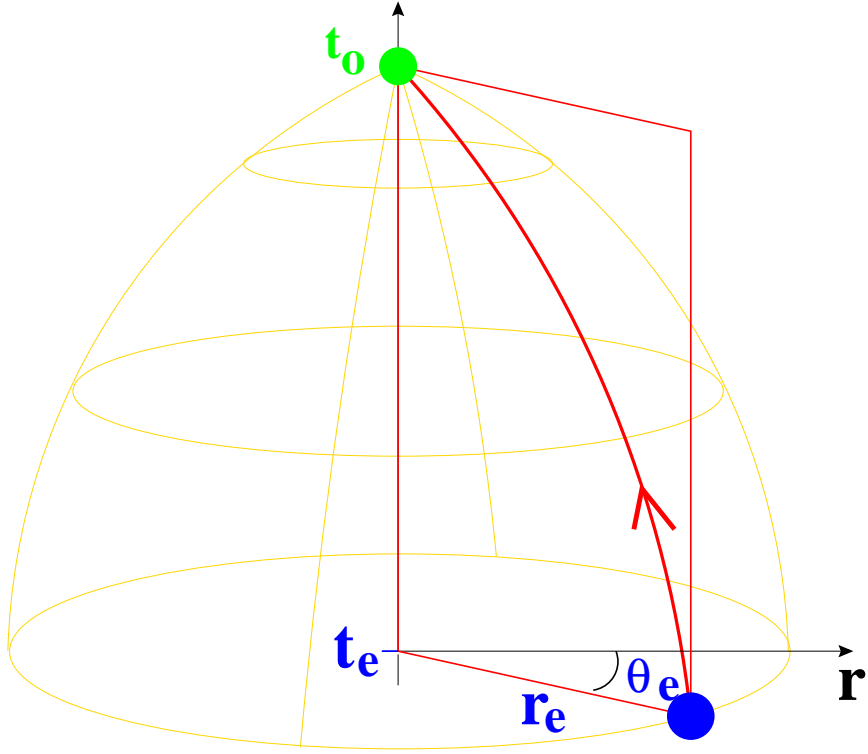}
\end{center}
\label{fig11}
\caption{An illustration of the propagation of photons in
our Universe. The dimensions shown here are $(t, r, \theta)$:
we skip $\phi$ for the purpose of representation. We are siting
at the origin, and at a time $t_0$, we can see a the light
of a galaxy emitted at $(t_e, r_e, \theta_e)$. Before
reaching us, the light from this galaxy has traveled over a 
curved trajectory.
In any point, the slope $dr/dt$ is given by
equation (\ref{photon.propagation}). So, the relation
between $r_e$ and $(t_0 - t_e)$ depends on the spatial curvature
and on the scale factor evolution. The
trajectory would be a straight line in space-time
only if $k=0$ and $a=constant$, i.e., in the limit of
Newtonian mechanics in Euclidean space.
The ensemble of all possible photon trajectories crossing $r=0$ at
$t=t_0$ is called our ``past light cone'', visible here in orange.
Asymptotically, near the origin, it can be approximated by 
a linear cone with $dl=cdt$, showing that at small distance, the physics
is approximately Newtonian.
}
\end{figure}

\vspace{0.5cm}

{\bf The redshift.}

First, a simple calculation based on equation (\ref{photon.propagation})
- we don't include it here - gives the redshift associated with 
a given source of light.
Take two observers sitting on two galaxies (with fixed comoving
coordinates). A light signal is sent from one observer to the other.
At the time of emission $t_1$, the first observer measures the wavelength
$\lambda_1$. At the time of reception $t_2$,
the second observer will measure the wavelength $\lambda_2$
such that
\begin{equation}
z= \frac{\delta \lambda}{\lambda} = 
\frac{\lambda_2-\lambda_1}{\lambda_1} = \frac{a(t_2)}{a(t_1)} -1~.
\label{redshift}
\end{equation}
So, the redshift depends on the variation of the scale-factor between
the time of emission and reception, but not on the curvature parameter
$k$.  This can be understood as if the scale-factor represented the
``stretching'' of light in out curved space-time. When the Universe
expands ($a(t_2) > a(t_1)$), the wavelengths are enhanced
($\lambda_2>\lambda_1$), and the spectral lines are red-shifted
(contraction would lead to blue-shift, i.e., to a negative redshift
parameter $z<0$).

In real life, what we see when we observe the sky is never
directly the distance, size, or velocity of an object, but the light
traveling from it. So, by studying spectral lines, we can easily
measure the redshifts.
This is why astronomers, when they refer to an object or to an event
in the Universe, generally mention the redshift rather than the
distance or the time. It's only
when the function $a(t)$ is known - and as we will see later, we
almost know it in our Universe - 
that a given redshift can be related to
a given time and distance.

The importance of the redshift as a measure of time and distance comes
from the fact that we don't see our full space--time, but only our
past line--cone, i.e., a three-dimensional subspace of the full
four--dimensional space--time, corresponding to all points which could
emit a light signal that we receive today on Earth. We can give a
representation of the past line--cone by removing the coordinate
$\phi$ (see figure 1.11). Of course, the cone is curved, like all
photon trajectories.

Also note that in Newtonian mechanics, the redshift was defined
as $z=v/c$, and seemed to be limited to $|z|<1$. The true GR expression
doesn't have such limitations, since the ratio of the scale factors
can be arbitrarily large without violating any fundamental principle.
And indeed, observations do show many objects - like
quasars - at redshifts of 2 or even bigger. We'll see later that we also
observe the Cosmic Microwave Background at a redshift of approximately 
$z=1000$!

\vspace{0.5cm}

{\bf The Hubble parameter in General Relativity.} 

In the limit of small redshift, we expect to recover the Newtonian
results, and to find a relation similar to $z=v/c=H r/c$. To show this,
let's assume that $t_0$ is the present time, and that our galaxy is 
at $r=0$. We want to compute the redshift of a nearby galaxy, which emitted 
the light that we receive today at a time $t_0-dt$. 
In the limit of small $dt$, the equation of propagation of light
shows that the physical distance $L$ between the galaxy and us is simply
\begin{equation}
L \simeq dl = c~ dt 
\end{equation}
while the redshift of the galaxy is
\begin{equation}
\label{def:hubble:GR}
z = \frac{a(t_0)}{a(t_0-dt)} - 1 
\simeq 
\frac{a(t_0)}{a(t_0)-\dot{a}(t_0)dt} - 1 
= \frac{1}{1-\frac{\dot{a}(t_0)}{a(t_0)}dt} - 1
\simeq
\frac{\dot{a}(t_0)}{a(t_0)} dt~.
\end{equation}
By combining these two relations we obtain
\begin{equation}
z \simeq \frac{\dot{a}(t_0)}{a(t_0)} \frac{L}{c} ~.
\end{equation}
So, at small redshift, we recover the Hubble law, and the role of the Hubble
parameter is played by $\dot{a}(t_0) / a(t_0)$. In the Friedmann Universe, 
we will directly define the Hubble parameter as the expansion rate of the 
scale factor:
\begin{equation}
H(t) =  \frac{\dot{a}(t)}{a(t)}~.
\end{equation}
The present value of $H$ is generally noted $H_0$:
\begin{equation}
H_0 = 100 \, h ~{\rm km~s}^{-1} {\rm Mpc}^{-1} , \qquad 0.5<h<0.8. 
\end{equation}

\vspace{0.5cm}

{\bf Angular diameter -- redshift relation.}

When looking at the sky, we don't see directly the size of the objects,
but only their angular diameter.
In Euclidean space, i.e. in absence of gravity, the angular
diameter $d \theta$ of an object would be related to its size $dl$ 
and distance $r$ through
\begin{equation}
d \theta = \frac{dl}{r} ~.
\end{equation}
Recalling that $z= v/c$ and $v = H r$, we easily find an angular diameter
-- redshift relation valid in Euclidean space:
\begin{equation}
\label{ang:diam:new}
d \theta = \frac{H ~dl}{c ~z}~.
\end{equation}
In General Relativity, because of the bending of light by gravity,
the steps of the calculation are different. Using the definition 
of infinitesimal distances (\ref{friedman.distance}), we see
that the physical size $dl$ of an object is related to its angular
diameter $d \theta$ through
\begin{equation}
d l = a(t_e) ~r_e ~d \theta
\label{angular.diameter}
\end{equation}
where $t_e$ is the time at which the galaxy emitted the light ray that 
we observe today on Earth, and $r_e$ 
is the comoving coordinate of the object.
The equation of motion of light gives a relation between $r_e$ and 
$t_e$:
\begin{equation}
\int_{r_e}^{0} \frac{- dr}{\sqrt{1-kr^2}} 
= \int_{t_e}^{t_0} \frac{c}{a(t)} dt
\label{integral}
\end{equation}
where $t_0$ is the time today. So, the relation between $r_e$ and 
$t_e$ depends on $a(t)$ and $k$. 
If we knew the function $a(t)$ and the value of $k$, 
we could integrate (\ref{integral}) explicitly and 
obtain some function $r_e(t_e)$. We would also know the relation
$t_e(z)$ between redshift and time of emission. So, we could
re-express equation
(\ref{angular.diameter}) as 
\begin{equation}
d \theta = \frac{dl}{a(t_e(z)) \,\, r_e(t_e(z))}~.
\end{equation}
This relation is called the angular diameter -- redshift
relation.  In the limit $z\ll1$, we get from (\ref{def:hubble:GR}) that
$z = H (t_0-t_e)$ and from (\ref{integral}) that $a_e r_e = c\,
(t_0-t_e)$, so we recover the Newtonian expression
(\ref{ang:diam:new}). Otherwise, we can define a function of redshift
$f(z)$ accounting for general relativistic corrections:
\begin{equation}
d \theta= \frac{H~dl}{c ~z} f(z)~, 
\qquad {\rm lim}_{z \rightarrow 0} f(z) = 1
\end{equation}
where $f$ depends on the dynamics of expansion and on the curvature.

A generic consequence is that
in the Friedmann Universe, for an object
of fixed size and redshift, the angular
diameter depends on the curvature - as illustrated graphically in figure
1.12.
\begin{figure}[!bt]
\begin{center}
\epsfysize=10cm
\epsfbox{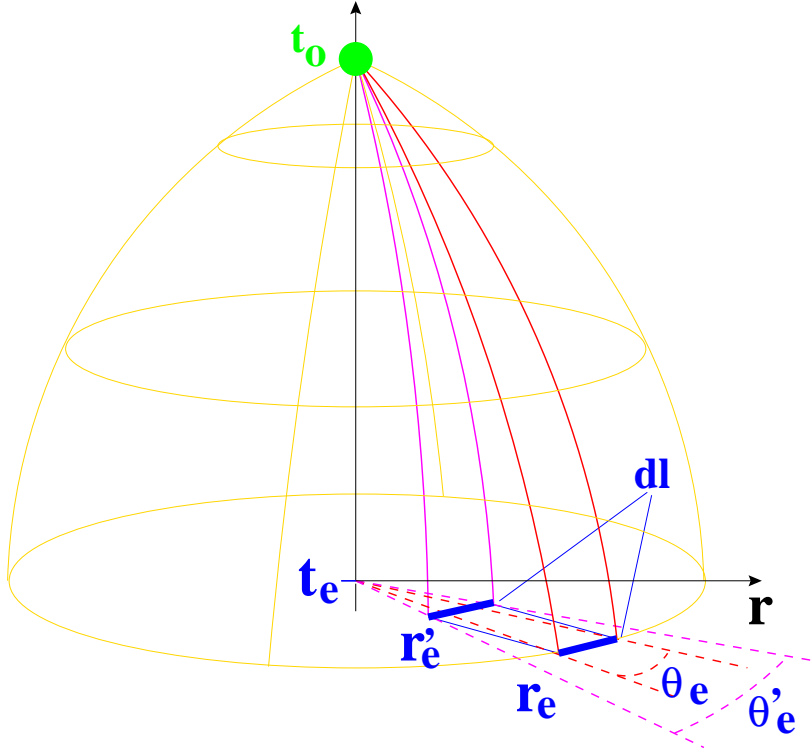}
\end{center}
\label{fig12}
\caption{
Angular diameter -- redshift relation. We consider an object
of fixed size $dl$ and fixed redshift, sending a light signal at time
$t_e$ that we receive at present time $t_0$. All photons travel
by definition with $\theta=$constant. However, the bending
of their trajectories in the $(t,r)$ plane depends on the
spatial curvature and on the scale factor evolution. So, for fixed $t_e$,
the comoving coordinate of the object, $r_e$, depends on curvature.
The red lines are supposed to illustrate the trajectory of light
in a flat Universe with $k=0$. If we keep $dl$, $a(t)$ and $t_e$ fixed,
but choose a positive value $k>0$, we know from equation (\ref{integral})
that the new coordinate ${r_e}'$ has to be smaller. But $dl$ is fixed,
so the new angle $d \theta'$ has to be bigger, 
as easily seen on the figure for
the purple lines. 
So, in a closed Universe, objects are seen under a larger angle.
Conversely, in an open Universe, they are seen under a smaller
angle.
}
\end{figure}
Therefore, if we know in advance the
physical size of an object,
we can simply measure its redshift, its angular diameter,
and immediately obtain some informations on the geometry of the Universe.

It seems very difficult to know in advance the physical size of a
remote object. However, we will see in the next chapter that some
beautiful developments of modern cosmology enable physicists to know
the physical size of the dominant anisotropies of the
Cosmological Microwave Background, visible at a redshift of $z \simeq
1000$. So, the angular diameter --
redshift relation has been used in the past decade in order to measure
the spatial curvature of the Universe. We will show the results in the
last section of chapter two.

\vspace{0.5cm}

{\bf Luminosity distance -- redshift relation.}

In Newtonian mechanics, the absolute
luminosity of an object and the apparent luminosity $l$ 
that we measure on Earth per unit of surface are related by
\begin{equation}
l = \frac{L}{4 \pi r^2} = \frac{L H^2}{4 \pi c^2 z^2} ~.
\label{Ll}
\end{equation}
So, if for some reason we know independently the absolute luminosity
of a celestial body (like for cepheids), and we measure its redshift,
we can obtain the value of $H$, as Hubble did in 1929.

But we would like to extend this technique to very distant objects (in
particular, supernovae of type Ia, which are observed up to a redshift
of two, and have a measurable absolute luminosity like cepheids).  For
this purpose, we need to compute the equivalent of (\ref{Ll}) in the
framework of general relativity. We define the {\it luminosity
distance} as
\begin{equation}
d_L \equiv \sqrt{\frac{L}{4 \pi l}}
\end{equation}
In absence of expansion, $d_L$ would simply correspond to the distance
to the source. On the other hand, in general relativity, it is easy to 
understand that equation (\ref{Ll}) is replaced by a more complicated
relation
\begin{equation}
\label{lum:dist}
l = \frac{L}{4 \pi \, a^2(t_0) \, r_e^2 (1+z)^2} 
\end{equation}
leading to
\begin{equation}
d_L = a(t_0) \, r_e (1+z) ~.
\end{equation}
Let us explain this result. First, the reason for the presence of the
factor $4 \pi \, a^2(t_0) \, r_e^2$ in equation (\ref{lum:dist}) is
obvious. The photons emitted at a comobile coordinate $r_e$ are
distributed today on a sphere of comobile radius $r_e$ surrounding the
source. Following the expression for infinitesimal distances
(\ref{friedman.distance}), the physical surface of this sphere is
obtained by integrating over $ds^2 = a^2(t_0) \, r_e^2 \sin \! \theta
\, d \theta \, d \phi$, which gives precisely $4 \pi \, a^2(t_0) \,
r_e^2$.  In addition, we should keep in mind that $L$ is a flux (i.e.,
an energy by unit of time) and $l$ a flux density (energy per unit of
time and surface). But the energy carried by each photon is inversely
proportional to its physical wavelength, and therefore to $a(t)$. This
implies that the energy of each photon has been divided by $(1+z)$
between the time of emission and now, and explains one of the two
factors $(1+z)$ in (\ref{lum:dist}).  The other factor comes from the
change in the rate at which photons are emitted and received. To see
this, let us think of one single period of oscillation of the
electromagnetic signal. If a wavecrest is emitted at time $t_e$ and
received at time $t_0$, while the following wavecrest is emitted at
time $t_e+\delta t_e$ and received at time $t_0+ \delta t_0$, we
obtain from the propagation of light equation (\ref{integral}) that:
\begin{equation}
\int_{t_e}^{t_0} \frac{c}{a(t)} dt
=
\int_{t_e+\delta t_e}^{t_0+\delta t_0} \frac{c}{a(t)} dt
=
\int_{r_e}^{0} \frac{- dr}{\sqrt{1-kr^2}} 
\end{equation}
If we concentrate on the first equality and rearrange the limits of 
integration, we obtain:
\begin{equation}
\int_{t_e}^{t_e+\delta t_e} \frac{c}{a(t)} dt
=
\int_{t_0}^{t_0+\delta t_0} \frac{c}{a(t)} dt~.
\end{equation}
Moreover, as long as the frequency of oscillation is very large with respect
to the expansion rate of the Universe (which means that
$a(t_e) \simeq a(t_e+\delta t_e)$ and
$a(t_0) \simeq a(t_0+\delta t_0)$), we can simplify this relation into:
\begin{equation}
\frac{\delta t_e}{a(t_e)}
=
\frac{\delta t_0}{a(t_0)}~.
\end{equation}
This indicates that the change between the rate of emission and reception
is given by a factor
\begin{equation}
\frac{(\delta t_0)^{-1}}{(\delta t_e)^{-1}}
=
\frac{a(t_e)}{a(t_0)}=(1+z)^{-1}~.
\end{equation}
This explains the second factor $(1+z)$ in (\ref{lum:dist}).

Like in the case of the angular diameter -- redshift relation, we would need
to know the function $a(t)$ and the value of $k$ 
in order to calculate explicitly the functions 
$r_e(t_e)$ and $t_e(z)$. Then, we would finally get a luminosity distance --
redshift relation, under the form
\begin{equation}
d_L = a(t_0) \,\, r_e(t_e(z)) \,\, (1+z) ~.
\end{equation}

If for several objects we can measure independently the 
absolute luminosity, the apparent luminosity and the redshift, we can
plot a luminosity distance versus redshift
diagram. For small redshifts $z \ll 1$, we should obtain a linear
relation, since in that case it is easy to see that at leading order
\begin{equation}
d_L \simeq a(t_e) \, r_e(t_e(z)) \simeq c z / H_0 ~.
\end{equation}
This is equivalent to plotting a Hubble diagram. However, at large
redshift, we should get a non-trivial curve whose shape would depend
on the spatial curvature and the dynamics of expansion.  We will see
in the next chapter that such an experiment has been performed for
many supernovae of type Ia, leading to one of the most intriguing
discovery of the past years.

\vspace{0.5cm}

In summary of this section, according to General Relativity, 
the homogeneous Universe is curved
by its own matter content, and the space--time
curvature can be described
by one number plus one function: the spatial curvature $k$, and 
the scale factor $a(t)$. We should be able to relate these two quantities
with the source of curvature: the matter density.

\subsection{The Friedmann law}
\label{sec136}

The Friedmann law relates the scale factor $a(t)$, the spatial
curvature parameter $k$ and the homogeneous energy density of the 
Universe $\rho(t)$:
\begin{equation}
\label{friedmann}
\left( \frac{\dot{a}}{a} \right)^2 
= \frac{8 \pi {\cal G}}{3} \frac{\rho}{c^2}
- \frac{k c^2}{a^2}~.
\end{equation}
Together with the propagation of light equation, this law
is the key ingredient of the Friedmann-Lema\^{\i}tre model.

In special/general relativity, the total energy of a particle 
is the sum of its rest energy $E_0=m c^2$ plus its momentum energy.
So, if we consider only non-relativistic particles like those forming
galaxies, we get $\rho = \rho_{\rm mass} c^2$. 
Then, the Friedmann equation looks exactly like
the Newtonian expansion law (\ref{newton.expansion}), excepted that
the function $r(t)$ (representing previously the position of 
objects)
is replaced by the scale factor $a(t)$. Of course, the two equations look
the same, but they are far from being equivalent. First, 
we have already seen in section \ref{sec135}
that although the distinction 
between the scale factor $a(t)$ and the classical position $r(t)$
is irrelevant at short distance, the difference of interpretation 
between the two is crucial at large
distances -- of order of the Hubble radius. Second, we have seen 
in section \ref{sec123}
that the
term proportional to $k$ seems to break the homogeneity of the Universe
in the Newtonian formalism, while in the Friedmann model, when it is 
correctly interpreted as the spatial curvature term, it is perfectly 
consistent with the Cosmological Principle.

Actually, there is a third crucial difference between the Friedmann
law and the Newtonian expansion law. In the previous paragraph, 
we only considered non-relativistic matter like galaxies.
But the Universe also contains relativistic particles 
traveling at the speed of light, like photons
(and also neutrinos if their mass is very small).
{\it A priori}, their gravitational effect on the
Universe expansion could be important. How can we include them?

\subsection{Relativistic matter and Cosmological constant}
\label{sec137}

The Friedmann equation is true for any types of matter, relativistic or
non-relativistic; if there are different
species, the total energy density $\rho$ is the sum over the density
of all species.

There is something specific about the type of matter considered: it is
the relation between $\rho(t)$ and $a(t)$, i.e., the rate at which 
the energy of a given fluid gets diluted by the expansion.

For non-relativistic matter, the answer is obvious. Take a 
distribution of particles with fixed comoving coordinates. 
Their energy density is given by their mass density 
times $c^2$. Look only at the matter contained into a
comoving sphere centered around the origin,
of comoving radius $r$. If the sphere is small with respect to the 
radius of spatial curvature, 
the physical volume inside the sphere is just $V = \frac{4 \pi}{3} (a(t) r)^3$.
Since both the sphere and the matter particles have fixed comoving coordinates,
no matter can enter or leave from inside the sphere during the expansion. 
Therefore, the mass (or the energy) inside the sphere is conserved. 
We conclude that $\rho V$ is constant and that $\rho \propto a^{-3}$.

For ultra--relativistic matter like photons, the energy of each particle
is not
given by the rest mass but by the frequency $\nu$
or the wavelength $\lambda$: 
\begin{equation}
E= h \nu = h c / \lambda .
\end{equation}
But we know that physical wavelengths are stretched proportionally
to the scale factor $a(t)$.
So, if we repeat the argument of the sphere, assuming now that it contains
a homogeneous bath of photons with equal wavelength, we see that the total 
energy inside the sphere evolves proportionally to $a^{-1}$. So, the energy
density of relativistic matter is proportional to $\rho \propto a^{-4}$.
If the scale factor increases, the photon energy density decreases faster
than that of ordinary matter.

This result could be obtained differently. For any gas of particles with
a given velocity distribution, 
one can define a pressure (corresponding 
physically to the fact that some particles would hit the borders 
if the gas was 
enclosed into a box). This can be extended to a gas of relativistic particles, 
for which the speed equals the speed of light. A calculation based on 
statistical mechanics gives the famous result that the pressure of a 
relativistic gas is related to its density by $p=\rho/3$.

In the case of the Friedmann universe, General Relativity provides several
equations: the Friedmann law, and also, for each species, an
equation of conservation:
\begin{equation}
\label{conservation}
\dot{\rho} = - 3 \frac{\dot{a}}{a} (\rho + p) ~.
\end{equation}
This is consistent with what we already said. For non-relativistic matter,
like galaxies, the pressure is negligible (like in a gas of still 
particles), and we get
\begin{equation}
\dot{\rho} = - 3 \frac{\dot{a}}{a} \rho 
\qquad \Rightarrow \qquad
\rho \propto a^{-3}.
\end{equation}
For relativistic matter like photons, we get
\begin{equation}
\dot{\rho} = - 3 \frac{\dot{a}}{a} (1 + \frac{1}{3}) \rho = 
- 4 \frac{\dot{a}}{a} \rho 
\qquad \Rightarrow \qquad 
\rho \propto a^{-4}.
\end{equation}
Finally, in quantum field theory, it is well--known that the vacuum
can have an energy density different from zero. The Universe could
also contain this type of energy (which can be related to particle
physics, phase transitions and spontaneous symmetry breaking). The
pressure of the vacuum is given by $p=-\rho$, in such way that the
vacuum energy is never diluted and its density
remains constant. This constant
energy density was called by Einstein -- who introduced it in a
completely different way and with other motivations -- the
Cosmological Constant. We will see that this term is probably
playing an important role in our Universe.

\chapter{The Standard Cosmological Model}
\label{chap2}

The real Universe is not homogeneous: it contains stars, galaxies,
clusters of galaxies... 

In cosmology, all quantities -- like the
density and pressure of each species -- are decomposed into a spatial
average, called the {\it background}, 
plus some inhomogeneities. The later are assumed to be small with
respect to the background in the early Universe: so, they can be
treated like linear perturbations. As a consequence, the Fourier modes
evolve independently from each other.
During the evolution, if the perturbations
of a given quantity become large, the linear approximation breaks down, and
one has to employ a full non--linear description, which is very
complicated in practice. 
However, for many purposes in cosmology, the
linear theory is sufficient in order to make testable predictions.

In section \ref{sec21}, we will describe the evolution of the
homogeneous background. Section \ref{sec22} will give some hints about
the evolution of linear perturbations -- and also, very briefly, about
the final non-linear evolution of matter perturbations.  Altogether,
these two sections provide a brief summary of what is called the
standard cosmological model, which depends on a few free
parameters. In section \ref{sec23}, we will show that the main
cosmological parameters have already been measured with quite good
precision. Finally, in section \ref{sec24}, we will introduce the
theory of inflation, which provides some initial conditions both for
the background and for the perturbations in the very early
Universe. We will conclude with a few words on the so-called
quintessence models.

\section{The Hot Big Bang scenario}
\label{sec21}

{\it A priori}, we don't know what type of fluid or particles
gives the dominant contributions to the
energy density of the Universe. According to the Friedmann
equation, this question is related to
many fundamental issues, like the behavior of
the scale factor, the spatial curvature, or the past and future
evolution of the Universe...

\subsection{Various possible scenarios for the history of the Universe}
\label{sec211}

We will classify the various types of matter that could fill the Universe 
according to their pressure-to-density ratio. The three most likely
possibilities are:

\begin{enumerate}
\item
ultra--relativistic particles, 
with $v \simeq c$, $~~p= \rho / 3$, $~~\rho \propto a^{-4}$.
This includes photons, massless neutrinos, and eventually
other particles that would have a very small mass and would be traveling 
at the speed of light. The generic name for this kind of matter, which
propagates like electromagnetic radiation, is precisely ``radiation''.
\item
non-relativistic pressureless matter -- in general, simply called
``matter'' by opposition to radiation -- with 
$v \ll c$, $~~p \simeq 0$, $~~\rho \propto a^{-3}$.
This applies essentially to all structures in the Universe: 
planets, stars, clouds of gas, or galaxies seen as a whole.
\item
a possible cosmological constant, with time--invariant energy density
and $p= - \rho$, that
might be related to the vacuum of the theory describing elementary particles,
or to something more mysterious. 
Whatever it is, we leave such a constant term as an open possibility.
Following the definition given by Einstein, what is actually 
called the ``cosmological constant'' ${\Lambda}$ is not the energy density
$\rho_{\Lambda}$, but the quantity
\begin{equation}
\Lambda = 8 \pi {\cal G} \rho_{\Lambda} / c^2
\end{equation}
which has the dimension of the inverse square of a time.
\end{enumerate}

We write the Friedmann equation including these three terms:
\begin{equation}
H^2 = \left( \frac{\dot{a}}{a} \right)^2 
= \frac{8 \pi {\cal G}}{3 c^2} \rho_{\rm R} ~+~
\frac{8 \pi {\cal G}}{3 c^2} \rho_{\rm M} 
~-~ \frac{k c^2}{a^2} 
~+~ \frac{\Lambda}{3} 
\end{equation}
where $\rho_{\rm R}$ is the radiation density and $\rho_{\rm M}$ the matter density.
The order in which we wrote the four terms on the right--hand side
-- radiation, matter, spatial curvature, cosmological constant -- 
is not arbitrary.
Indeed, they evolve with respect to the scale factor as $a^{-4}$, $a^{-3}$, 
$a^{-2}$ and $a^{0}$. 
So, if the scale factors keeps growing, and if
these four terms are present in the Universe,
there is a chance that they all dominate the expansion of the Universe 
one after each other (see figure 2.1). 
\begin{figure}[!bt]
\begin{center}
\epsfxsize=12cm
\epsfbox{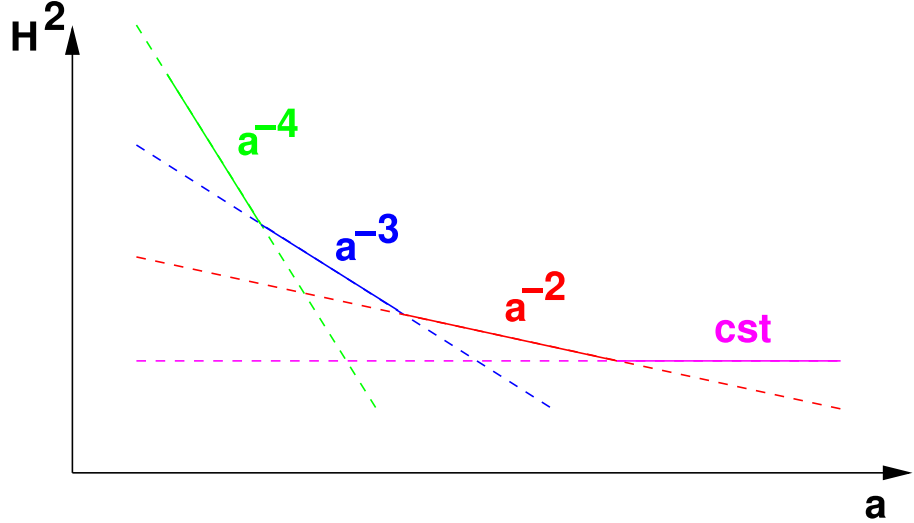}
\end{center}
\caption{Evolution of the square of the Hubble parameter, 
in a scenario in which all typical contributions to the Universe expansion 
(radiation, matter, curvature, cosmological constant)
dominate one after each other.
}
\end{figure}
Of course,
it is also possible that some of these terms do not exist at all, or 
are simply negligible.
For instance, some possible scenarios would be:
\begin{itemize}
\item
only matter domination, from the initial singularity until today
(we'll come back to the notion of
Big Bang later).
\item
radiation domination $\rightarrow$ matter domination today.
\item
radiation dom. $\rightarrow$  matter dom.
$\rightarrow$ curvature dom. today
\item
radiation dom. $\rightarrow$ matter dom.
$\rightarrow$ cosmological constant dom. today
\end{itemize}
But all the cases that do not respect the order (like for instance:
curvature domination $\rightarrow$ matter domination) are impossible.

During each stage, one component strongly dominates the others, and the
behavior of the scale factor, of the Hubble parameter and 
of the Hubble radius are given by:
\begin{enumerate}
\item Radiation domination: 
\begin{equation}
\frac{\dot{a}^2}{a^2} \propto a^{-4}, 
\qquad
a(t) \propto t^{1/2},
\qquad
H(t) = \frac{1}{2 t}, 
\qquad
R_H(t) = 2 c t.
\end{equation}
So, the Universe is in decelerated power--law expansion.
\item Matter domination: 
\begin{equation}
\frac{\dot{a}^2}{a^2} \propto a^{-3}, 
\qquad
a(t) \propto t^{2/3},
\qquad
H(t) = \frac{2}{3 t}, 
\qquad
R_H(t) = \frac{3}{2} c t.
\end{equation}
Again, the Universe is in power--law expansion, but it decelerates
more slowly than during radiation domination.
\item Negative curvature domination ($k < 0$):
\begin{equation}
\frac{\dot{a}^2}{a^2} \propto a^{-2}, 
\qquad
a(t) \propto t,
\qquad
H(t) = \frac{1}{t}, 
\qquad
R_H(t) = c t.
\end{equation}
An open Universe dominated by its curvature is in linear expansion.
\item Positive curvature domination: if $k>0$, and if there is no cosmological
constant, the right--hand side 
finally goes to zero: expansion stops. After, 
the scale factor starts to decrease. $H$ is negative, but
the right--hand side of the Friedmann equation remains positive. The Universe 
recollapses. We know that we are not in such a phase, because
we observe the Universe expansion. But {\it a priori},
we might be living in a closed Universe, slightly before the expansion stops.

\item Cosmological constant domination:
\begin{equation}
\frac{\dot{a}^2}{a^2} \rightarrow {\rm constant}, 
\qquad
a(t) \propto \exp(\Lambda t / 3),
\qquad
H = c / R_H = \sqrt{\Lambda /3}.
\end{equation}
The Universe ends up in exponentially accelerated expansion.
\end{enumerate}

So, in all cases, there seems to be a time in the past at which the
scale factor goes to zero, called the initial singularity or the ``Big
Bang''. The Friedmann description of the Universe is not supposed to
hold until $a(t)=0$. At some time, when the density reaches a critical
value called the Planck density, we believe that gravity has to be
described by a quantum theory, where the classical notion of time and
space disappears. Some proposals for such theories exist, mainly in
the framework of ``string theories''. Sometimes, string theorists try
to address the initial singularity problem, and to build various
scenarios for the origin of the Universe. Anyway, this field is still
very speculative, and of course, our understanding of the origin of
the Universe will always break down at some point. A reasonable goal
is just to go back as far as possible, on the basis of testable
theories.

The future evolution of the Universe heavily depends on the existence
of a cosmological constant. If the later is exactly zero, then the
future evolution is dictated by the curvature (if $k>0$, the Universe
will end up with a ``Big Crunch'', where quantum gravity will show up
again, and if $k \leq 0$ there will be eternal decelerated expansion).
If instead there is a positive cosmological term which never decays
into matter or radiation, then the Universe necessarily ends up in
eternal accelerated expansion.

\subsection{The matter budget today}
\label{sec212}

In order to know the past and future evolution of the Universe, it
would be enough to measure the present density of radiation, matter
and $\Lambda$, and also to measure $H_0$. Then, thanks to the
Friedmann equation, it would be possible to extrapolate $a(t)$ at any
time\footnote{At least, this is true under the simplifying assumption
that one component of one type does not decay into a component of
another type: such decay processes actually take place in the early
universe, and could possibly take place in the future.}. Let us
express this idea mathematically.  We take the Friedmann equation,
evaluated today, and divide it by $H_0^2$:
\begin{equation}
1 = \frac{8 \pi {\cal G}}{3 H_0^2 c^2} 
\left(\rho_{{\rm R} 0} + \rho_{{\rm M}0} \right) 
- \frac{k c^2}{a_0^2 H_0^2} + \frac{\Lambda}{3 H_0^2} .
\end{equation}
where the subscript $0$ means ``evaluated today''.
Since by construction, the sum of these four terms is one,
they represent the relative contributions
to the present Universe expansion. These terms are usually written
\begin{eqnarray}
\Omega_{\rm R} &=& \frac{8 \pi {\cal G}}{3 H_0^2 c^2} \rho_{{\rm R} 0}, \\
\Omega_{\rm M} &=& \frac{8 \pi {\cal G}}{3 H_0^2 c^2} \rho_{{\rm M} 0}, \\
\Omega_k &=& \frac{k c^2}{a_0^2 H_0^2}, \\
\Omega_{\Lambda} &=& \frac{\Lambda}{3 H_0^2}, \\
\end{eqnarray}
and the ``matter budget'' equation is
\begin{equation}
\Omega_{\rm R} + \Omega_{\rm M} - \Omega_k + \Omega_{\Lambda} = 1.
\end{equation}
The Universe is flat provided that
\begin{equation}
\Omega_0 \equiv \Omega_{\rm R} + \Omega_{\rm M} + \Omega_{\Lambda}
\end{equation}
is equal to one. In that case, as we already know, the total density
of matter, radiation and $\Lambda$ is equal at any time
to the critical density
\begin{equation}
\rho_c(t) = \frac{3 c^2 H^2(t)}{8 \pi {\cal G}}.
\end{equation}
Note that the parameters $\Omega_x$, where $x \in \{ R, M,\Lambda \}$, 
could have been defined as the present
density of each species divided by the 
present critical density:
\begin{equation}
\Omega_x = \frac{\rho_{x0}}{\rho_{c0}}.
\end{equation}
So far, we conclude that the evolution of the Friedmann Universe can be 
described entirely in terms of four parameters, called the
``cosmological parameters'':
\begin{equation}
\Omega_{\rm R} , \Omega_{\rm M}, \Omega_{\Lambda}, H_0. 
\end{equation}
One of the main purposes of observational cosmology is to measure the
value of these cosmological parameters.

\subsection{The Cold and Hot Big Bang alternatives}
\label{sec213}

Curiously, after the discovery of the Hubble expansion and of the
Friedmann law, there were no significant progresses in cosmology for a
few decades.  The most likely explanation is that most physicists were
not considering seriously the possibility of studying the Universe in
the far past, near the initial singularity, because they thought that
it would always be impossible to test any cosmological model
experimentally.

Nevertheless, a few pioneers tried to think about the origin of the
Universe. At the beginning, for simplicity, they assumed that
the expansion of the Universe was always dominated by a single
component, the one forming galaxies, i.e., pressureless matter. Since
going back in time, the density of matter increases as $a^{-3}$,
matter had to be very dense at early times. This was
formulated as the ``Cold Big Bang'' scenario.  

According to Cold Big Bang, in the early Universe, the density was so
high that matter had to consist in a gas of nucleons and electrons.
Then, when the density fell below a critical value, some nuclear
reactions formed the first nuclei - this era was called
nucleosynthesis. But later, due to the expansion, the dilution of
matter was such that nuclear reactions were suppressed (in general, the
expansion freezes out all processes whose characteristic
time--scale becomes smaller than the so--called Hubble time--scale
$H^{-1}$).  So, only the
lightest nuclei had time to form in a significant amount.  After
nucleosynthesis, matter consisted in a gas of nuclei and electrons, with
electromagnetic interactions. When the density become even smaller,
they finally combined into atoms -- this second transition is called the
recombination. At late time, any small density
inhomogeneity in the gas of atoms was enhanced by gravitational
interactions. The atoms started to accumulate into clumps like stars and
planets - but this is a different story.

In the middle of the XX-th century,
a few particle physicists tried to build the first models of
nucleosynthesis -- the era of nuclei formation. In particular, four
groups -- each of them not being aware of the work of the others --
reached approximately the same negative conclusion: in the Cold
Big Bang scenario, nucleosynthesis does not work properly, because the
formation of hydrogen is strongly suppressed with respect to that of
heavier elements. But this conclusion is at odds with observations:
using spectrometry, astronomers
know that there is a lot of hydrogen in
stars and clouds of gas. The groups of the Russo-American Gamow in the
1940's, of the Russian Zel'dovitch (1964), of the British Hoyle and
Tayler (1964), and of Peebles in Princeton (1965) all reached this
conclusion.  They also proposed a possible way to reconcile
nucleosynthesis with observations. If one assumes that during
nucleosynthesis, the dominant energy density is that of photons, the
expansion is driven by $\rho_{\rm R} \propto a^{-4}$, and the rate of
expansion is different. This affects the kinematics of the nuclear
reactions in such way that enough hydrogen can be created.

In that case, the Universe would be
described by a Hot Big Bang scenario, in which the radiation density
dominated at early time. Before nucleosynthesis and recombination, 
the mean free path of the photons was very small,
because they were continuously interacting -- first, with electrons and
nucleons, and then, with electrons and nuclei. So, their motion could
be compared with the Brownian motion in a gas of particles: they formed
what is called a ``black--body''. 
In any black--body, the many interactions maintain
the photons in thermal equilibrium, and their spectrum 
(i.e., the number density of photons as a function of wavelength)
obeys to a law found by Planck in the 1890's. Any ``Planck spectrum''
is associated with a given temperature.

Following the Hot Big Bang scenario, after recombination, the
photons did not see any more charged electrons and nuclei, but only 
neutral atoms. So, they stopped interacting significantly with matter.
Their mean free path became infinite, and they simply traveled along
geodesics -- excepted a very small fraction of them which interacted
accidentally with atoms, but since matter got diluted, this
phenomenon remained subdominant.  So, essentially, the photons
traveled freely from recombination until now, keeping the same energy
spectrum as they had before, i.e., a Planck spectrum, but with a
temperature that decreased with the expansion. This is an effect of
General Relativity: the wavelength of an individual photon is
proportional to the scale factor; so the shape of the Planck spectrum is 
conserved, but the whole spectrum is shifted in wavelength.
The temperature of a black--body
is related to the energy of an average photon with average wavelength:
$T \sim <\! E \!> \sim h c / < \! \lambda \! >$. So, the temperature
decreases like $1/ < \! \lambda \! >$, i.e., like $a^{-1}(t)$. 
\begin{figure}[!bt]
\begin{center}
\epsfxsize=12cm
\epsfbox{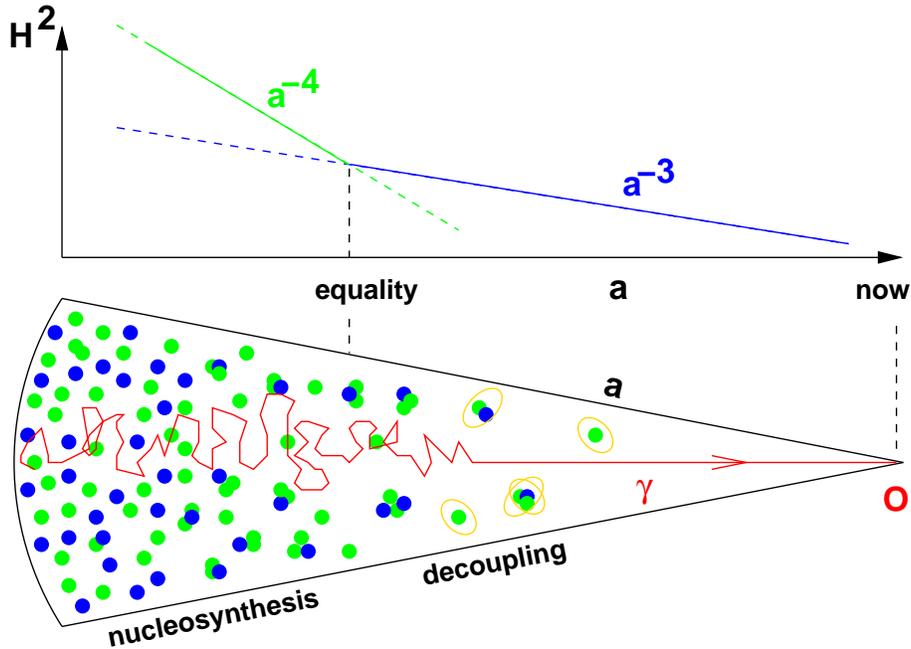}
\end{center}
\caption{On the top, evolution of the square of the Hubble parameter 
as a function
of the scale factor in the Hot Big Bang scenario. We see the two stages
of radiation and matter domination. On the bottom, an idealization of
a typical photon trajectory. Before decoupling, the mean free path is
very small due to the many interactions with baryons and electrons. 
After decoupling,
the Universe becomes transparent, and the photon travels in straight line,
indifferent to the surrounding distribution of electrically neutral matter.
}
\end{figure}

The physicists that we mentioned above noticed that these photons
could still be observable today, in the form of a homogeneous
background radiation with a Planck spectrum.  Following their
calculations -- based on nucleosynthesis -- the present temperature of
this cosmological black--body had to be around a few Kelvin
degrees. This would correspond to typical wavelengths of the order of one
millimeter, like microwaves.

\subsection{The discovery of the Cosmic Microwave Background}
\label{sec214}

These ideas concerning the Hot Big Bang scenario remained completely unknown, 
excepted from a small number of theorists.

In 1964, two American radio--astronomers, A. Penzias and R. Wilson,
decided to use a radio antenna of unprecedented sensitivity -- 
built initially for telecommunications -- in order to
make some radio observations of the Milky Way. They discovered
a background signal, of equal intensity in all directions, 
that they attributed to instrumental noise. However,
all their attempts to eliminate this noise failed.

By chance, it happened that Penzias phoned to a friend at MIT, Bernard
Burke, for some unrelated reason. Luckily, Burke asked about the 
progresses of the experiment. But Burke had recently spoken with
one of his colleagues, Ken Turner, who was just back from a visit
Princeton, during which he had followed a seminar by Peebles about
nucleosynthesis and possible relic radiation. Through this 
series of coincidences, Burke could put Penzias in contact with the
Princeton group. After various checks, it became clear that
Penzias and Wilson had made the first measurement of a 
homogeneous radiation with a Planck spectrum and a temperature
close to 3~Kelvins:
the Cosmic Microwave Background (CMB). Today, the CMB temperature
has been measured with great precision: $T_0=2.726$~K.

This fantastic observation was a very strong evidence in favor of
the Hot Big Bang scenario. It was also the first
time that a cosmological model was checked experimentally.
So, after this discovery,
more and more physicists realized that reconstructing the detailed
history of the Universe was not purely science fiction, and started to
work in the field.

The CMB can be seen in our everyday life: fortunately,
it is not as powerful as a microwave oven, but when we look at 
the background noise on the screen of a TV set, one fourth of the power
comes from the CMB!

\subsection{The Thermal history of the Universe}
\label{sec215}

Between the 1960's and today, a lot of efforts have been made in order
to study the various stages of the Hot Big Bang scenario with
increasing precision. Today, some models about given epochs in the
history of Universe are believed to be well understood, and are
confirmed by observations; some others remain very speculative.

For the earliest stages, there are still many competing scenarios --
depending, for instance, on assumptions concerning string theory.
Following the most conventional picture, gravity became a classical
theory (with well--defined time and space dimensions) at a time called
the Planck time
\footnote{By convention, the origin of time is chosen by extrapolating
the scale-factor to $a(0)=0$. Of course, this is only a convention, it
has no physical meaning.  }: $t \sim 10^{-36}$s, $\rho \sim (10^{18}
{\rm GeV})^4$.  Then, there was a stage of inflation (see section
\ref{sec24}), possibly related to GUT (Grand Unified Theory) symmetry
breaking at $t \sim 10^{-32}$s, $\rho \sim (10^{16} {\rm GeV})^4$.
After, during a stage called reheating, the scalar field
responsible for inflation decayed into a thermal bath of
standard model particles like
quarks, leptons, gauge bosons and Higgs bosons. The EW (electroweak)
symmetry breaking occured at $t \sim 10^{-6}$s, $\rho \sim (100~ {\rm
GeV})^4$.  Then, at $t \sim 10^{-4}$s, $\rho \sim (100~ {\rm MeV})^4$,
the quarks combined themselves into hadrons (baryons and mesons).

After these stages, the Universe entered into a series of processes
that are much better understood, and well
constrained by observations. These are:
\begin{enumerate}
\item at $t=1 - 100$~s, $T=10^9 - 10^{10}$~K, $\rho \sim (0.1 - 1
~{\rm MeV})^4$, a stage called nucleosynthesis, responsible for the
formation of light nuclei, in particular hydrogen, helium and
lithium. By comparing the theoretical predictions with the observed
abundance of light elements in the present Universe, it is possible to
give a very precise estimate of the total density of baryons in the
Universe: $\Omega_{\rm B} h^2=0.021 \pm 0.005$.
\item at $t\sim 10^4$~yr, $T\sim 10^4$~K, 
$\rho \sim (1 ~{\rm eV})^4$, the radiation density
equals the matter density: the Universe goes from radiation domination
to matter domination. 
\item at $t\sim 10^5$~yr, $T\sim 2500$~K, $\rho \sim (0.1 ~{\rm
eV})^4$, the recombination of atoms causes the decoupling of
photons. After that time, the Universe is almost transparent: the photons
free--stream along geodesics. So, by looking at the CMB, we obtain a
picture of the Universe at decoupling. In first approximation,
nothing has changed in the
distribution of photons between $10^{5}$~yr and today, excepted for an
overall redshift of all wavelengths (implying $\rho \propto a^{-4}$,
and $T \propto a^{-1}$).
\item after recombination, the small inhomogeneities
of the smooth matter distribution are amplified. This leads to the
formation of stars and galaxies, as we shall see later.
\end{enumerate}

The success of the Hot big bang Scenario relies on the existence of a
radiation--dominated stage followed by a matter--dominated stage. However, an
additional stage of curvature or cosmological constant 
domination is not excluded.

\subsection{A recent stage of curvature or cosmological constant domination?}
\label{sec216}

If today, there was a very large contribution of the curvature and/or
cosmological constant to the expansion ($\Omega_{\rm M} \ll |\Omega_k|$ or
$\Omega_{\rm M} \ll |\Omega_{\Lambda}|$), the deviations from the Hot Big Bang
scenario would be very strong and incompatible with many observations.
However, nothing forbids that
$\Omega_k$ and/or $\Omega_{\Lambda}$ are of order one. In that case,
they would play a significant part in the Universe expansion only 
in a recent epoch (typically, starting at a redshift of one or two). 
Then, the main predictions of the conventional 
Hot Big Bang scenario would be only slightly modified.
For instance, we could live in a close or open Universe with
$|\Omega_k| \sim 0.5$: in that case, there would be a huge radius of 
curvature, with observable consequences only at very large distances.

In both cases $\Omega_k = {\cal O}(1)$ and $\Omega_{\Lambda} = {\cal
O}(1)$, the recent evolution of $a(t)$ would be affected (as clearly
seen from the Friedmann equation). So, there would be a modification
of the luminosity distance--redshift relation, and of the angular
diameter--redshift relation.  We will see in section \ref{sec23} how
this has been used in recent experiments.

Another consequence of a recent $k$ or $\Lambda$ domination would
be to change the age of the Universe. Our knowledge of
the age of the Universe comes from the measurement of the Hubble 
parameter today. Within a given cosmological scenario, it is possible to
calculate the function $H(t)$
(we recall that by convention, the origin of time is 
such that $a \rightarrow 0$ when $t \rightarrow 0$).
Then, by measuring the present value $H_0$, we obtain immediately the
age of the Universe $t_0$.
The result does not depend very much on what happens
in the early Universe. Indeed, equality takes place very early, at $10^4$yr.
So, when we try to evaluate roughly the age of the Universe in billions of 
years, we are essentially sensitive to what happens {\it after} equality.

In a matter--dominated Universe, $H = 2 / (3t)$. Then, using the
current estimate of $H_0$, we get $t_0 = 2/ (3 H_0) \sim 9$~Gyr.  If
we introduce a negative curvature or a positive cosmological constant,
it is easy to show (using the Friedmann equation) that $H$ decreases
more slowly in the recent epoch. In that case, for the same value of
$H_0$, the age of the Universe is obviously bigger.  For instance,
with $\Omega_{\Lambda} \simeq 0.7$, we would get 
$t_0 \sim 14$~Gyr.

We leave the answer to these intriguing issues for section \ref{sec23}.

\subsection{Dark Matter}
\label{sec217}

We have many reasons to believe that the non-relativistic 
matter is of two kinds: ordinary matter, and dark matter.
One of the well-known evidences for dark matter arises from 
galaxy rotation curves.

Inside galaxies, the stars orbit around the center. If we can measure 
the redshift
in different points inside a given galaxy, we can reconstruct the
distribution of velocity $v(r)$ as a function of the distance $r$
to the center.
It is also possible to measure the distribution of luminosity $I(r)$
in the same galaxy. What is not directly observable is the 
mass distribution $\rho(r)$.
However, it is reasonable to assume that the mass distribution
of the {\it observed luminous matter} is proportional to the
luminosity distribution: $\rho_{\rm lum}(r) = b ~I(r)$, 
where $b$ is an unknown coefficient of
proportionality called the bias. 
From this, we can compute the gravitational potential $\Phi_{\rm lum}$
generated by the luminous matter, and the corresponding orbital
velocity, given by ordinary Newtonian mechanics:
\begin{eqnarray}
\rho_{\rm lum}(r) &=& b~ I(r), \\
\Delta \Phi_{\rm lum} (r) &=& 4 \pi {\cal G} ~ \rho_{\rm lum}(r), \\
v^2_{\rm lum} (r) &=& r \frac{\partial}{\partial r} \Phi_{\rm lum} (r).
\end{eqnarray}
So, $v_{\rm lum} (r)$ is known up to an arbitrary normalization factor
$\sqrt{b}$. However, for many galaxies, even by varying $b$, it is impossible
to obtain a rough agreement between $v(r)$ and $v_{\rm lum}(r)$
(see figure 2.3). 
The stars rotate faster than expected at large radius. We conclude that
there is some non--luminous matter, which deepens the potential well
of the galaxy. 
\begin{figure}[!bt]
\begin{center}
\epsfxsize=12cm
\epsfbox{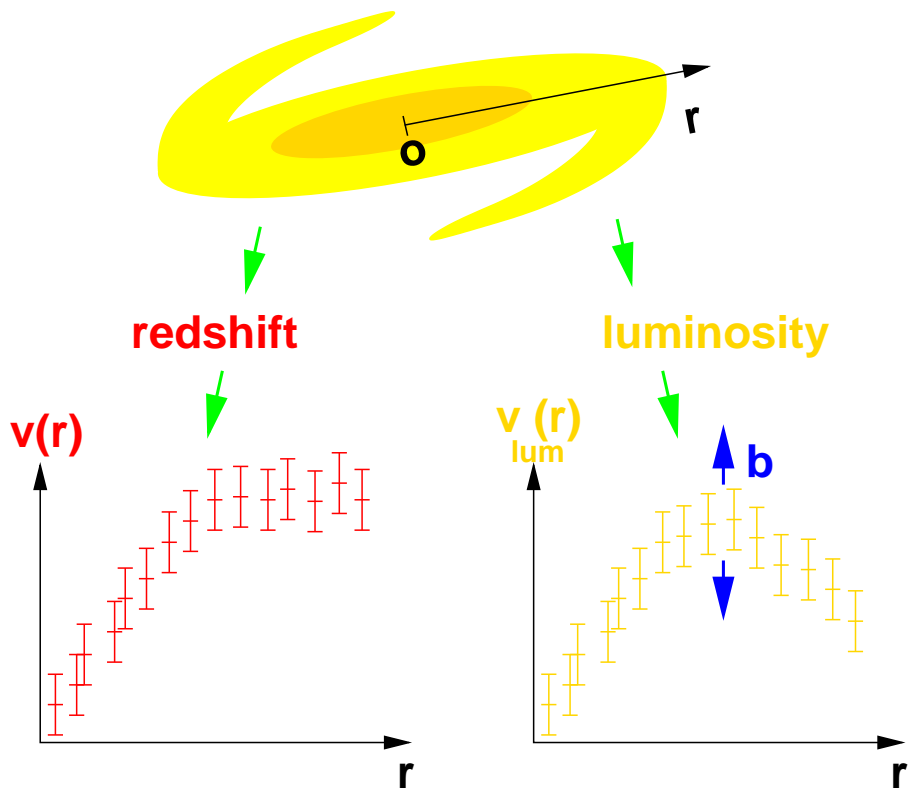}
\end{center}
\caption{A sketchy view of the galaxy rotation curve issue. 
The genuine orbital velocity of the stars is measured
directly from the redshift. From the luminosity distribution, we can reconstruct
the orbital velocity under the assumption that all the mass in the galaxy
arises form of the observed luminous matter. Even by varying the unknown 
normalization parameter $b$, it is impossible to obtain an agreement between 
the two curves: their shapes are different, with the reconstructed velocity
decreasing faster with $r$ than the genuine velocity. So, there
has to be some non--luminous matter around, deepening the potential 
well of the galaxy.
}
\end{figure}

Apart from galactic rotation curves, there are many arguments
-- of more cosmological nature -- which imply the presence of a
large amount of non--luminous matter in the Universe, called dark matter.
For various reasons, it cannot consist in
ordinary matter that would remain invisible just
because it is not lighten up. Dark matter has to be composed of particle
that are intrinsically uncoupled with photons -- unlike ordinary matter, 
made up of baryons. Within the standard model of particle
physics, a good candidate for non-baryonic dark matter
would be a neutrino with a small mass. Then, 
dark matter would be relativistic (this hypothesis is called
Hot Dark Matter or HDM).
However, HDM is excluded by some types of observations:
dark matter particles have to be non-relativistic, otherwise
galaxy cannot form during matter domination. Non--relativistic dark matter
is generally called Cold Dark Matter (CDM). 

There are a few candidates for CDM in various extensions of the standard 
model of particle physics: for instance, some supersymmetric partners of gauge 
bosons (like the neutralino or the gravitino), or the axion of the
Peccei-Quinn symmetry. Despite many efforts, these particles have never
been observed directly in the laboratory. This is not completely
surprising, given that they are -- by definition -- very weakly coupled
to ordinary particles.

In the following, we will decompose $\Omega_{\rm M}$ in 
$\Omega_{\rm B} + \Omega_{\rm CDM}$. This  introduces one more cosmological
parameter. With this last ingredient, we have described the main features
of the Standard Cosmological Model, at the level of homogeneous
quantities. We will now focus on the perturbations of this background.


\section{Cosmological perturbations}
\label{sec22}

\subsection{Linear perturbation theory}
\label{sec221}

All quantities, such as densities, pressures, velocities, curvature,
etc., can be decomposed into a spatial average plus some inhomogeneities:
\begin{equation}
\rho_x(t, \vec{r}) = \bar{\rho}_x(t) + \delta \rho_x(t, \vec{r}).
\end{equation}
We know that the CMB temperature is approximately the same in all
directions in the sky. This proves that in the early Universe, at
least until a redshift $z \sim 1000$, the distribution of matter was
very homogeneous. So, we can treat the inhomogeneities as small linear
perturbations with respect to the homogeneous background. For a given
quantity, the evolution becomes eventually non--linear when the
relative density inhomogeneity
\begin{equation}
\delta_x(t, \vec{r}) = \frac{\delta \rho_x (t, \vec{r})}{\bar{\rho}_x (t)} 
\end{equation}
becomes of order one.

In linear perturbation theory, it is very useful to make a Fourier 
transformation, because the evolution of each Fourier mode is independent
of the others:
\begin{equation}
\delta_x(t, \vec{k}) = \int d^3 \! \vec{r} ~e^{-i \vec{k}.\vec{r}}
\delta_x(t, \vec{r}).
\end{equation}
Note that the Fourier transformation is defined with respect
to the comoving coordinates $\vec{r}$. So, $k$ is the
{\it comoving} wavenumber, and $2 \pi / k$ the
{\it comoving} wavelength. The {\it physical} wavelength is
\begin{equation}
\lambda(t) = \frac{2 \pi a(t)}{k}.
\end{equation}
This way of performing the Fourier transformation is the most
convenient one, leading to independent evolution for each mode. It
accounts automatically for the expansion of the Universe. If no
specific process occurs, $\delta_x(t, \vec{k})$ will remain constant,
but each wavelength $\lambda(t)$ is nevertheless stretched
proportionally to the scale factor.

Of course, to a certain extent, the perturbations in the Universe
are randomly distributed. 
The purpose of the cosmological
perturbation theory is not to predict the {\it individual}
value of each $\delta_x(t, \vec{k})$ in each direction, 
but just the {\it spectrum} of the perturbations at
a given time, i.e., the root--mean--square of all
$\delta_x(t, \vec{k})$ for a given time and
wavenumber, averaged over all directions.
This spectrum will be noted simply as $\delta_x(t, k)$.

The full linear theory of cosmological perturbation is quite
sophisticated. Here, we will simplify considerably. The most
important density perturbations to follow
during the evolution of the Universe are those of
photons, baryons, CDM, and finally, the space--time
curvature perturbations.
As far as space--time curvature is concerned,
the homogeneous background is fully described
by the Friedmann expression for infinitesimal distances, i.e.,
by the scale factor $a(t)$
and the spatial curvature parameter $k$.
We need to define some perturbations around this average
homogeneous curvature. The rigorous way to do it is rather
technical, but at the level of this course, the reader can admit that
the most relevant curvature perturbation is a simple function of
time and space, which is almost equivalent to
the usual Newtonian gravitational potential $\Phi (t, \vec{r})$
(in particular, at small distance, the two are completely equivalent).

\subsection{The horizon}
\label{sec222}

Now that we have seen which type of perturbations need to be studied,
let us try to understand their evolution. In practice, this amounts
in integrating a complicated system
of coupled differential equations, for each comoving Fourier mode $k$. 
But if we just want to understand
{\it qualitatively} the evolution of a given mode, 
the most important is to compare its
wavelength $\lambda(t)$ with a characteristic scale called the
{\it horizon}. Let us define this new quantity.

Suppose that at a time $t_1$, two photons are emitted in opposite
directions. The horizon $d_H$ at time $t_2$ relative to time $t_1$ is
defined as the physical distance between the two photons at time
$t_2$ (see figure 2.4).
\begin{figure}[!bt]
\begin{center}
\epsfxsize=12cm
\epsfbox{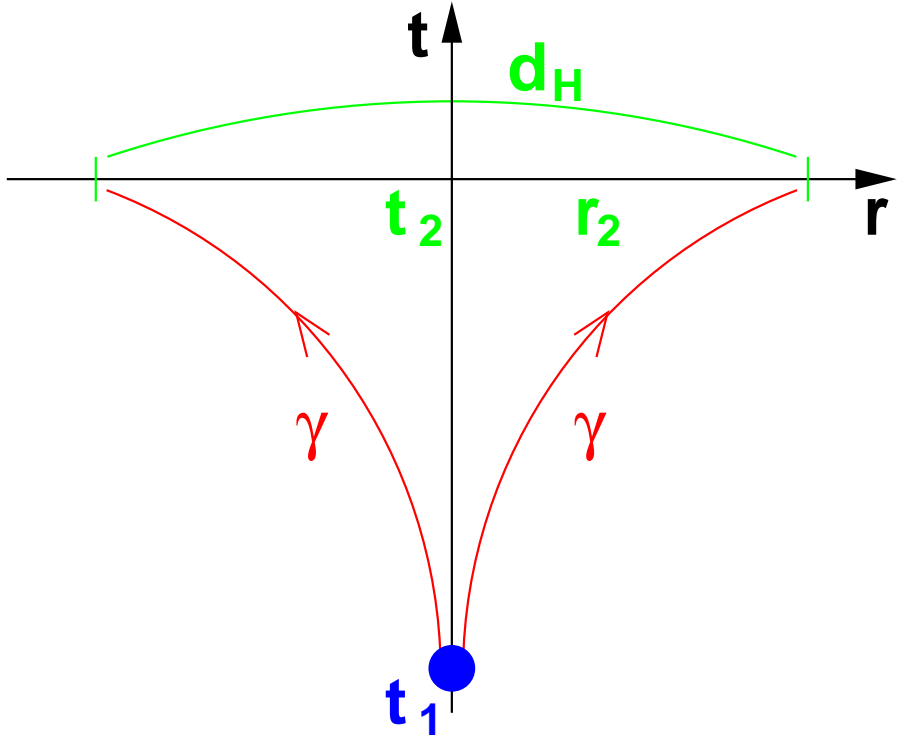}
\end{center}
\caption{
The horizon $d_H (t_1, t_2) $ is the physical distance at time
$t_2$ between two photons emitted in opposite directions at time $t_1$.
}
\end{figure}
In order to compute it, we note that if the
origin $r=0$ coincides with the point of emission, the 
horizon can be expressed as
\begin{equation}
d_H (t_1, t_2) = 2 \int_{0}^{r_2} dl
= 2 \int_{0}^{r_2} a(t_2) \frac{dr}{\sqrt{1 - k r^2}}.
\end{equation}
But the equation of propagation of light, applied to one of the photons,
also gives
\begin{equation}
\int_{t_1}^{t_2} \frac{c~dt}{a(t)} = 
\int_{0}^{r_2} \frac{dr}{\sqrt{1 - k r^2}}.
\end{equation}
Combining the two equations, we get
\begin{equation}
d_H (t_1, t_2) =
2 a(t_2) \int_{t_1}^{t_2} \frac{c~dt}{a(t)}.
\end{equation}
Physically, since no information can travel faster than light,
the horizon represents the maximal scale on which a process
starting at time $t_1$ can have an impact at time $t_2$. 
In the limit in which $t_1$ is very close to the initial
singularity, $d_H$ represents the {\it causal horizon},
i.e. the maximal distance at which two points in the Universe can
be in causal contact.
In particular, perturbations with wavelengths
$\lambda(t_2) > d_H(t_1, t_2)$ are {\it acausal}, and cannot
be affected by any physical process -- if they were, it would
mean that some information had traveled faster than light.

\vspace{0.5cm}

During {\bf radiation domination}, when $a(t) \propto t^{1/2}$,
it is straightforward to compute
\begin{equation}
d_H (t_1, t_2) =
4 ~c ~t_2^{1/2} [t_2^{1/2} - t_1^{1/2}].
\end{equation}
In the limit in which $t_1 \ll t_2$, we find
\begin{equation}
d_H (t_1, t_2) \rightarrow
4 ~c ~t_2 = 2 ~R_H(t_2).
\end{equation}
So, during radiation domination, the causal horizon coincides
with the Hubble radius.

\vspace{0.5cm}

During {\bf matter domination}, when $a(t) \propto t^{2/3}$,
we get
\begin{equation}
d_H (t_1, t_2) =
6~c ~t_2^{2/3} [t_2^{1/3} - t_1^{1/3}].
\end{equation}
When $ t_1 \ll t_2$ this becomes
\begin{equation}
d_H (t_1, t_2)
\rightarrow
6 ~c ~t_2 = 4 ~R_H(t_2).
\end{equation}
So, even during a matter--dominated stage following
a radiation--dominated stage, the causal horizon remains
of the same order of magnitude as the Hubble radius.

\vspace{0.5cm}

Since each wavelength $\lambda(t)=2 \pi a(t) /k$ 
evolves with time at a different rate than the Hubble radius $R_H(t)$, 
a perturbation of fixed wavenumber $k$ can be
super--horizon ($\lambda \gg R_H$) at some initial
time, and sub--horizon ($\lambda \ll R_H$) at a later time.
This is illustrated on figure 2.5. While the scale factor
is always decelerating, the Hubble radius grows linearly: so,
the Fourier modes enter one after each other inside the causal
horizon. The modes which enter earlier are those
with the smallest wavelength.
\begin{figure}[!bt]
\begin{center}
\epsfxsize=12cm
\epsfbox{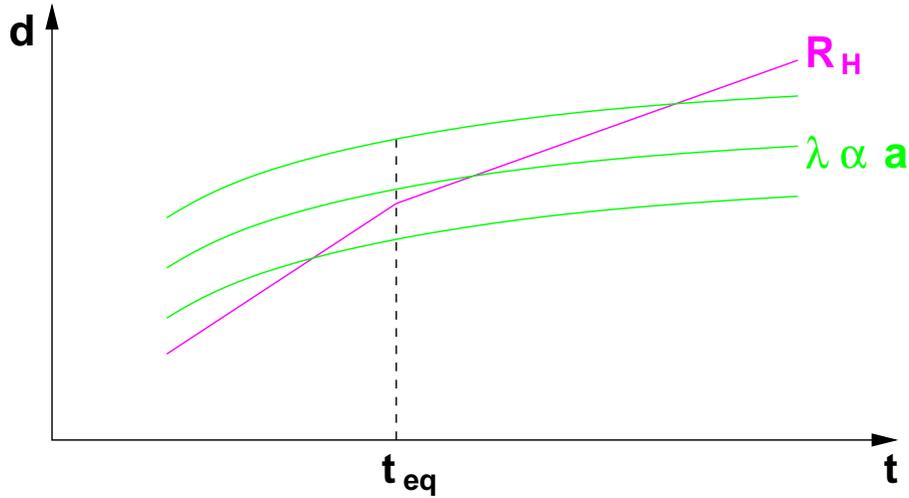}
\end{center}
\caption{
The wavelengths of some cosmological perturbations compared
with the Hubble radius, during radiation domination and matter domination.
Because all wavelengths
$\lambda(t)= 2 \pi a(t) / k$ grow with negative acceleration,
and $R_H(t)$ grows linearly, the modes enter one after each other
inside the horizon.
The smallest perturbations enter during radiation domination, the others
during matter domination. The largest cosmological perturbation
observable today (in CMB observations)
has $\lambda(t_0) = R_H(t_0)$.
}
\end{figure}

What about the perturbations that we are able to measure today?
Can they be super--horizon? In order to answer this question, we first note
that the largest distance that we can see today in our past--light--cone 
is the one between two photons emitted at decoupling,
and reaching us today from opposite direction in the sky. Because
the Universe was not transparent to light before decoupling, with current 
technology, we cannot think of any observation probing a larger
distance.

So, in order to compute the maximal wavelength accessible to
observation, we can make a short calculation, quite similar to that
of the horizon, but with reversed photon trajectories: 
the two photons are emitted at $(t_1, r_1)$, and they reach us today
at $(t_0, 0)$ from opposite directions. The propagation--of--light
equation gives
\begin{equation}
\int_{t_1}^{t_0} \frac{c~dt}{a(t)} = 
\int_{r_1}^{0} \frac{- dr}{\sqrt{1 - k r^2}}.
\end{equation}
But today, two points located at coordinate $r_1$
in opposite directions are separated by the physical distance
\begin{equation}
2 \int_0^{r_1} dl = 2 \int_0^{r_1} a(t_0) \frac{dr}{\sqrt{1 - k r^2}}
= 2 a(t_0) \int_{t_1}^{t_0} \frac{c~dt}{a(t)} = d_H(t_1, t_0).
\end{equation}
We conclude that the causal horizon evaluated today gives an upper
bound on the distances accessible to any observation (whatever is the
precise value of the time of decoupling). Our knowledge of the
Universe is limited to a finite sphere of physical radius $R_H(t_0)$.
So, all observable cosmological perturbations are sub--horizon today,
while in the past, they were super--horizon (see figure 2.5).  The
largest observable scale has entered inside the horizon very recently,
with a wavelength approximately equal to
\begin{equation}
R_H(t_0) = \frac{c}{H_0} \simeq \frac{3 \times 10^8 ~{\rm m~s}^{-1}}
{100~h ~{\rm km~s}^{-1}{\rm Mpc}^{-1}}
\simeq 3\times 10^3  ~h^{-1} {\rm Mpc}.
\end{equation}

\vspace{0.5cm}

All these general considerations will allow us
to understand qualitatively the evolution of the different 
perturbations, and especially those of photons and CDM, 
which are the most interesting.
We will try to visualize the evolution
on a three-dimensional diagram, like the one of figure
2.6. This representation gives
the perturbation amplitude $\delta_x (t,k)$
for a given species $x$. In other words, it gives the evolution
of each Fourier mode with respect to time. The purple line
shows the limit between causal (sub--horizon) perturbations on the
right, and acausal (super--horizon) perturbations on the left.
\begin{figure}[!bt]
\begin{center}
\epsfxsize=12cm
\epsfbox{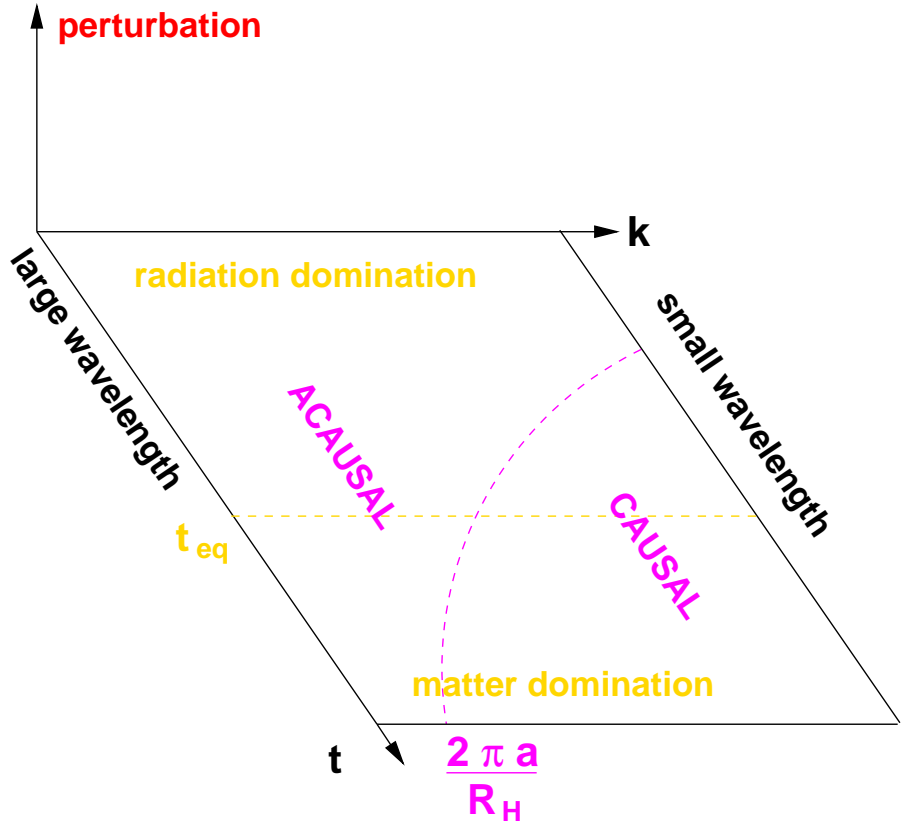}
\end{center}
\caption{
For any species, this type of diagram can be used for understanding
qualitatively
the evolution of the Fourier spectrum as a function of time.
Each Fourier mode of fixed comoving wavenumber $k$ enters into
the causal horizon when it crosses the purple line
(which corresponds to $\lambda(t) \sim R_H(t)$, i.e., 
to $k \sim \frac{2 \pi}{c} \, a(t) \, H(t)$).
}
\end{figure}

\subsection{Photon perturbations}
\label{sec223}

In good approximation, the
photons keep a Planck spectrum during all the relevant evolution. 
As a consequence, the photon density
perturbations $\delta_{\gamma}(t, k)$
can be reformulated as some temperature fluctuations
around the mean black--body temperature: $\delta T / \bar{T}(t, k)$
(where $\bar{T}$ is the spatial average of the temperature at time $t$).
The evolution is very different before and
after the time of photon decoupling. 

Indeed, before decoupling,
photons are strongly coupled to baryons through electromagnetic
interactions, and have a vanishing mean--free--path.  As a
consequence, the baryons and the photons are almost
indistinguishable: they share
the same density perturbations, and can be viewed as single fluid.
In addition, since baryons have a mass, they feel gravitational
forces: so, the entire baryon--photon fluid is coupled with
the gravitational potential.

So, in each well of gravitational potential, the
baryon--photon fluid tends to be compressed by gravity, but the
relativistic photon pressure $p=\frac{1}{3} \rho$ tends to
resist to this compression. The dynamics of the
fluid is governed by two antagonist forces, as for
any oscillator. So, for each Fourier mode, any departure from 
equilibrium -- such as
a non--zero initial perturbation -- will lead to oscillations. 
These oscillations propagate through the fluid exactly like
sound waves in the air: due to this analogy, the
perturbations in the baryon--photon fluid before decoupling are
called {\it acoustic oscillations}.

Of course, the oscillations can occur only for causal wavelengths:
because of causality, wavelengths bigger than the Hubble radius do
not feel any force. So, a given mode with
a given non--zero initial amplitude starts oscillating
only when it enters into the horizon (see figure 2.7). The
initial amplitude for each mode at initial time is given by
the {\it primordial spectrum} of perturbations. Of course,
if the Universe was perfectly homogeneous from the very beginning
(with no initial perturbations),
there would be no acoustic oscillations -- and no formation of
structure either, so we {\it must} assume some
primordial perturbations. At this stage, the
primordial spectrum has to be assumed arbitrarily.
Only in the last section, we will see a precise mechanism called 
``inflation''
{\it leading}
to a given primordial spectrum. On figure 2.7, we start from
a flat spectrum, independent of $k$ -- in good
approximation, this corresponds to what is
{\it observed} today, and also to what is {\it predicted} by the
theory of inflation. 
\begin{figure}[!bt]
\begin{center}
\epsfxsize=12cm
\epsfbox{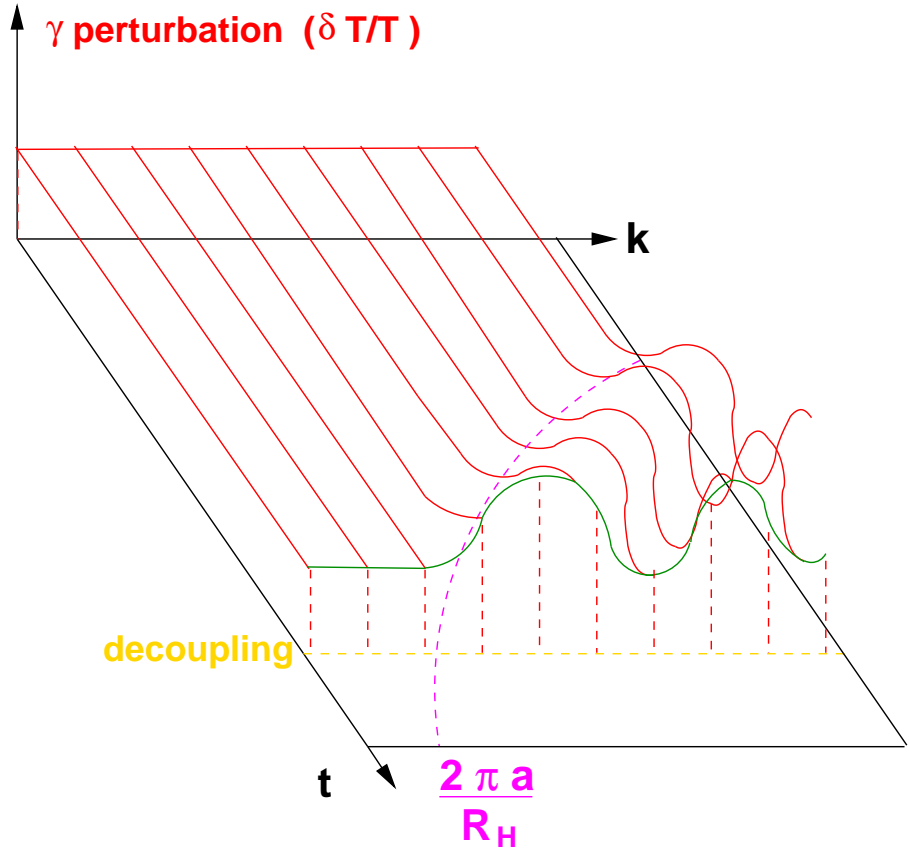}
\end{center}
\caption{
Evolution of the Fourier spectrum of photon
temperature fluctuations, as a function of time. We start from
a flat primordial spectrum, as predicted by inflation and as confirmed by 
observations. When a given Fourier mode enters inside the horizon
(i.e., crosses the dashed purple line),
it starts to oscillate. At decoupling, the temperature distribution
is frozen (later, all temperatures are redshifted by a factor
$1/a(t)$, so the relative fluctuation $\delta T/\bar{T}$ remains constant).
So, the Fourier spectrum measured today is the same as at the time
of decoupling. It exhibits a flat plateau for
wavelengths bigger than the Hubble radius at decoupling, and then
a series of peaks called {\it acoustic peaks}.
}
\end{figure}

When the photons decouple from the matter, their mean--free--path
grows rapidly from zero to infinity, and they travel in straight line
with essentially no interaction. 
During this {\it free-streaming}, all their characteristics --
momentum, energy, wavelength, spectrum, temperature -- do not evolve,
excepted under the effect of homogeneous expansion. So, $\delta T$
and $\bar{T}$ are both shifted proportionally to $1/a(t)$, but 
the distribution of $\delta T / \bar{T}(t, k)$ 
observed today is exactly the same as at the time of
decoupling. Observing
the temperature anisotropies in the sky amounts exactly in
taking a picture of the Universe at the time of 
decoupling\footnote{This statement is a bit simplified. In fact, there
are a few second-order effects altering the temperature
spectrum between decoupling and now, like the late reionization of the 
Universe, the gravitational lensing of CMB photons by nearby clusters,
or the decay of the gravitational potential at small redshift when the
Universe starts to accelerate...}.

So, in order to know the present Fourier spectrum of temperature
fluctuations, it is enough to follow the
acoustic oscillations {\it until decoupling}. As easily understood from
figure 2.7, the Fourier modes which are still outside the
Hubble radius at decoupling have no time to oscillate. Smaller
wavelengths have enough time for one oscillation, 
even smaller ones for two oscillations, etc. 
The Fourier spectrum of temperature fluctuations
consists in a flat plateau for 
$k < 2 \pi a(t_{\rm dec}) / R_H(t_{\rm dec})$, and then a first peak,
a second peak, a third peak, etc.

Note that according to this model, the
perturbations oscillate, but they keep the same order
of magnitude (there is no large amplification).
Therefore, if we start from small
primordial perturbations, we are able to describe all the evolution
of temperature fluctuations with the {\it linear} theory of perturbations.

\subsection{Observing the CMB anisotropies}
\label{sec224}

The measurement of the CMB by Penzias and Wilson showed no evidence
for perturbations -- in fact, we should say anisotropies, because we
can only map the temperature along different {\it directions}.
However, some CMB anisotropies must exist, because our Universe is not
homogeneous today.  The current inhomogeneities have to grow from
fluctuations in the past, and at the time of decoupling there must
have been already at least some small fluctuations in the densities,
and therefore, in the temperature.  The order of magnitude of CMB
anisotropies was predicted many years before being measured.  By
extrapolating from the present inhomogeneous structure back to the
time of decoupling, many cosmologists in the 80's expected $\delta T/
\bar{T} $ to be at least of order $10^{-6}$ -- otherwise, clusters of
galaxies could not have formed today.

Many experiments were devoted to the detection of these
anisotropies. The first successful one was COBE-DMR, an 
American satellite carrying an
interferometer of exquisite sensitivity. In 1992,
COBE mapped the anisotropies all over the sky,
and found an average amplitude $\delta T/ \bar{T} \sim 10^{-5}$
(see figure 2.8). This
was in perfect agreement with the theoretical predictions -- another
big success for cosmology. 
\begin{figure}[!th]
\begin{center}
\epsfxsize=12cm
\epsfbox{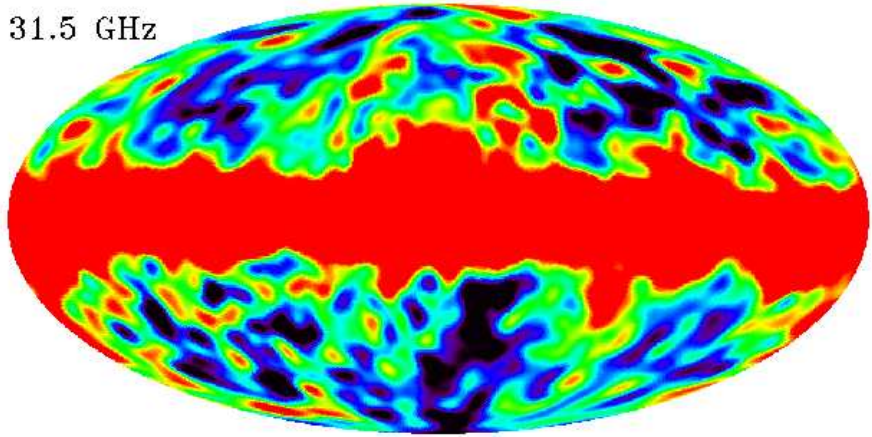}
\end{center}
\caption{ The first genuine ``picture of the Universe'' at the time of
decoupling, 100~000~years after the initial singularity, and
15~billion years before the present epoch. 
Each blue (resp. red) spot
corresponds to a slightly colder (resp. warmer) region of the Universe
at that time. This map, obtained by the American satellite COBE in
1994 (see C.~L.~Bennett et al., Astrophys.J. 464 (1996) L1-L4), 
covers the entire sky: so, it pictures a huge sphere centered on
us (on the picture, the sphere has been projected onto an ellipse, where
the upper and lower points represent the direction of the
poles of the Milky way). Away from the central red stripe, which corresponds
to photons emitted from our own galaxy,
the fluctuations are only of order $10^{-5}$ with respect to the
average value $T_0=2.728$~K. They are the ``seeds'' for the present
structure of the Universe: each red spot corresponds to a small
overdensity of photons and baryons at the time of decoupling, that has
been enhanced later, leading to galaxies and clusters of galaxies
today.  }
\end{figure}
The COBE experiment had an angular resolution of a few degrees:
so, anisotropies seen under one degree or less were smoothed
by the detector. In a Fourier decomposition, it means that COBE
could only measure the spectrum of wavelengths larger than
the Hubble radius at decoupling. So, it was not probing the
acoustic oscillations, but only the flat plateau.

COBE was the first experimental evidence in favor of the linear theory of 
cosmological perturbations.
The next step was to increase the angular resolution,
in order to probe smaller wavelengths,
check the existence of the acoustic peaks, and compare them with the theory.
In the 90's, there were some considerable efforts in this direction.

First, on the theoretical side, the spectrum of CMB anisotropies
was investigated with greater and greater precision. Some people
developed numerical codes computing the exact spectrum
for any choice of cosmological parameters. 
Indeed, the details of the peaks depend on many quantities.
For instance, the amplitude of the
acoustic oscillations is governed by the density of
baryons and photons. The whole final shape of the spectrum depends on 
the primordial spectrum of perturbations.
Last but not least, the position of the peaks depends on the Hubble radius at
decoupling. Fortunately, the theory is able to predict the
{\it physical size} of $R_H(t_{\rm dec})$, and the redshift
of decoupling -- so, by measuring the angular size
associated to the wavelength of the acoustic peaks, it should
be possible to use the angular diameter--redshift relation
(see section \ref{sec135}),
and to measure the {\it curvature} of the Universe!
We conclude that if the acoustic peaks could be measured,
the comparison with theoretical predictions would reveal the values of
the most crucial cosmological parameters.

On the experimental side, there were many successful experiments after COBE.
Particulary decisive progresses were made with
Boomerang, a US--Italian--Canadian balloon, carrying some detectors called
bolometers. In 2001,
Boomerang published the map of figure 2.9. It focuses on a small
patch of the sky, but with much better resolution than COBE
(a few arc--minutes).
\begin{figure}[!th]
\begin{center}
\epsfxsize=12cm
\epsfbox{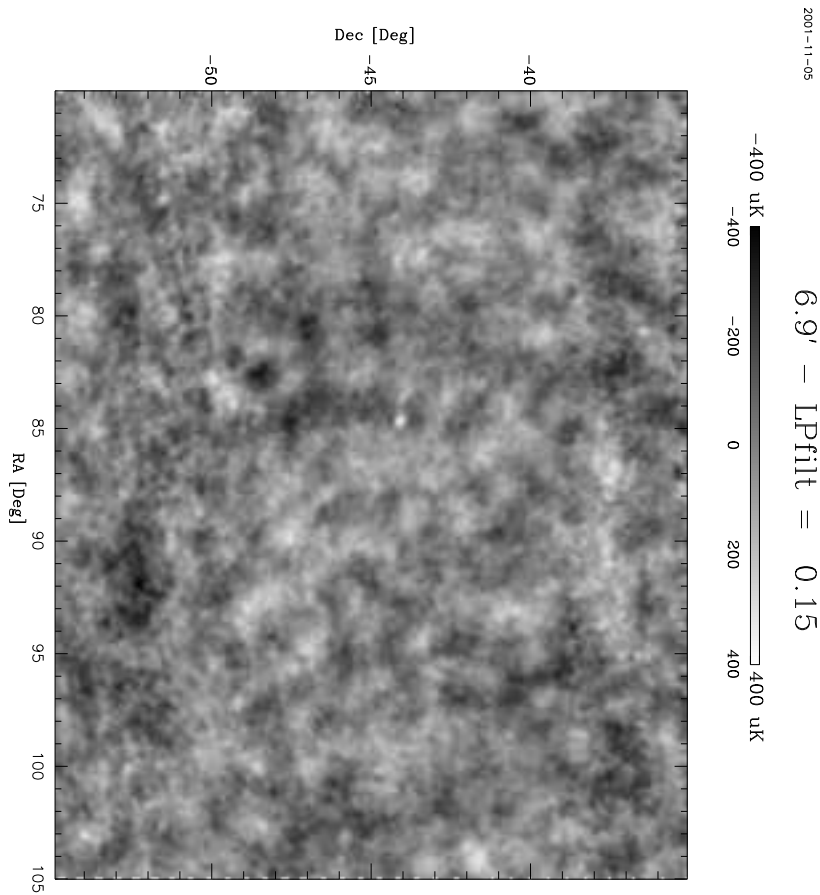}
\end{center}
\caption{
The map of CMB anisotropies obtained by the balloon experiment
Boomerang in 2001 (see S.~Masi et al., Prog.Part.Nucl.Phys. 48 (2002) 243-261).
Unlike COBE, Boomerang only analyzed a small patch of the sky, but
with a much better angular resolution of a few arc--minutes. The dark
(resp. light) spots correspond to colder
(resp. warmer) regions.
}
\end{figure}
The Fourier decomposition of the Boomerang map clearly shows
the first three acoustic peaks (see figure 2.10). The data points
are perfectly fitted with a theoretical curve obtained numerically,
for a simple choice of cosmological parameters (figure 2.11).

At the beginning of 2003, the NASA satellite WMAP published
a full-sky CMB map (see figure 2.12); it is the second one after that of COBE,
but with a resolution increased by a factor 30. 
The corresponding Fourier
spectrum is extremely precise on the scale of the first and
second acoustic peaks, and perfectly compatible with the theory 
(see figure 2.13). This beautiful
agreement with some predictions made several decades before the
experiment is one of the major successes of cosmology.
By fitting theoretical curves to this data, it is possible to
measure various cosmological parameters with already a good precision.
For instance, the position of the first peak is strongly in favor of $k=0$,
i.e., a spatially flat Universe. We will come back to these measurements 
in section \ref{sec23}. 
\begin{figure}[!th]
\begin{center}
\epsfxsize=12cm
\epsfbox{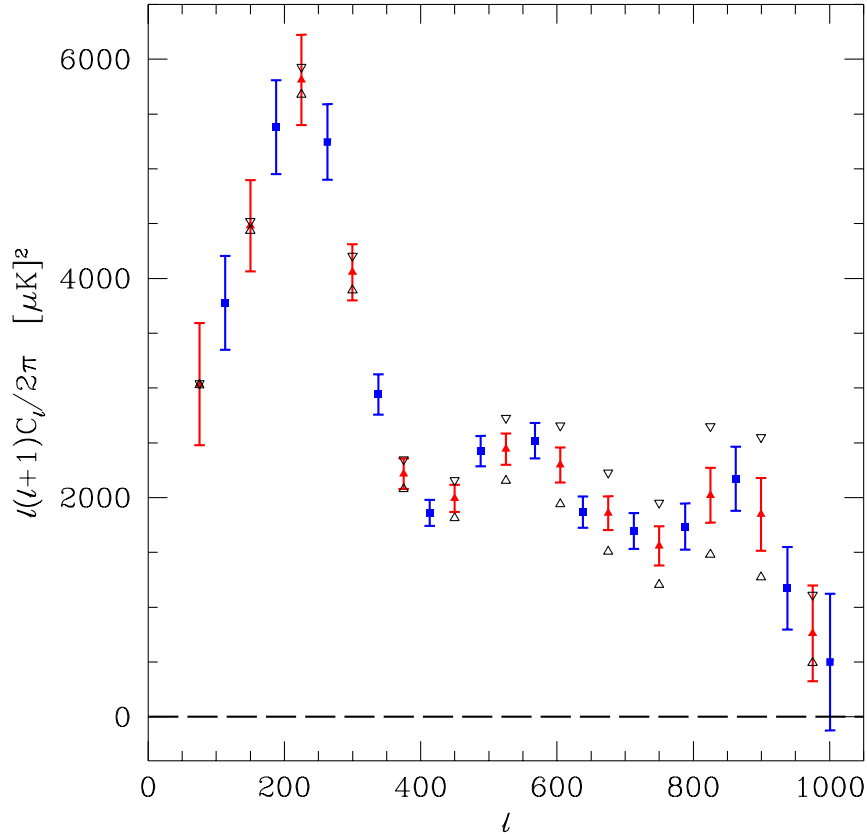}
\end{center}
\caption{
The Fourier decomposition of the Boomerang map reveals the structure
of the first three acoustic oscillations (see  C.~B.~Netterfield et al.,
Astrophys.J. 571 (2002) 604-614). These data points account
for the spectrum of $(\delta T)^2$ in ($\mu$K)$^2$ -- we know from
COBE that
$\delta T$ is roughly of order $10^{-5} T_0 \simeq
30 ~\mu{\rm K}$. The spectrum is not shown as a function
of wavenumber $k$, but of multipole number $l$. This is because the 
anisotropies
are not observed in 3-dimensional space, but on a two-dimensional sphere:
the sphere of redshift $z \sim 1000$ centered on us. For purely
geometric reasons, a spherical
map is not expandable in Fourier modes, but in Legendre multipoles.
However, the interpretation is the same. The first, second and third peak
correspond to the modes that could oscillate one, two or three times
between their entry into the Hubble radius and the time of decoupling.
}
\end{figure}
\begin{figure}[!th]
\begin{center}
\epsfxsize=12cm
\epsfbox{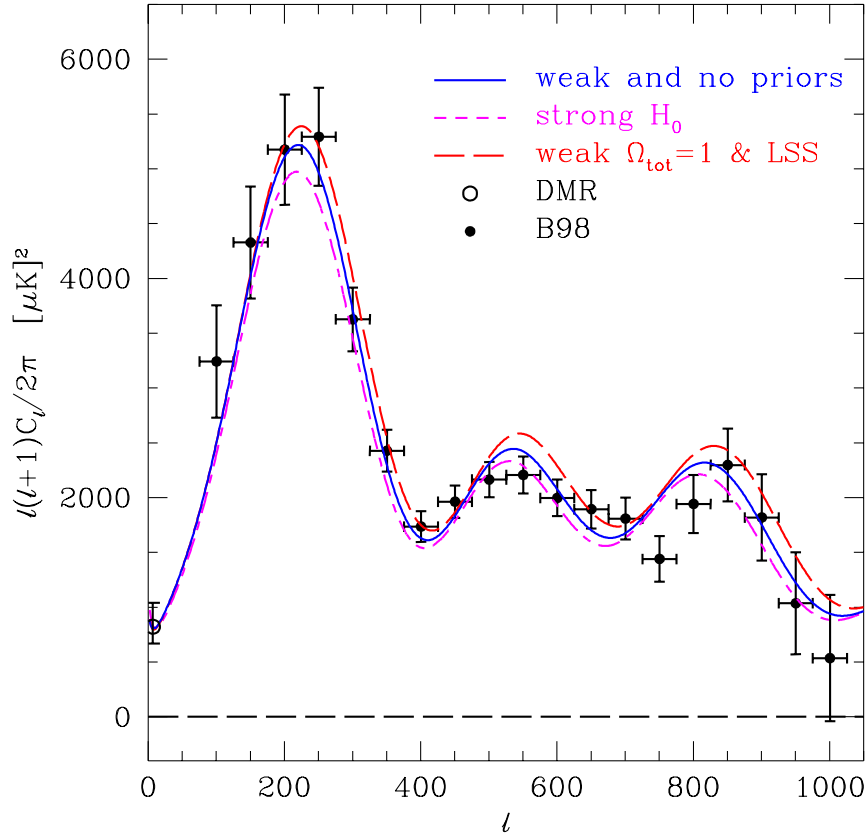}
\end{center}
\caption{ The spectrum of the Boomerang map is perfectly fitted by the
theoretical predictions (see  C.~B.~Netterfield et al.,
Astrophys.J. 571 (2002) 604-614). For various choices of cosmological
parameters -- such as $\Omega_{\rm B}$, $\Omega_{\rm CDM}$,
$\Omega_{\Lambda}$, $\Omega_{\rm R}$, $H_0$, the parameters describing
the primordial spectrum, etc.-- the theoretical calculation is
performed by a numerical code, leading to the curves shown here for
three different models. By comparing with the experimental data, one
can extract the parameter values, as we will see in section 2.3.  }
\end{figure}
\begin{figure}[!th]
\begin{center}
\epsfxsize=12cm
\epsfbox{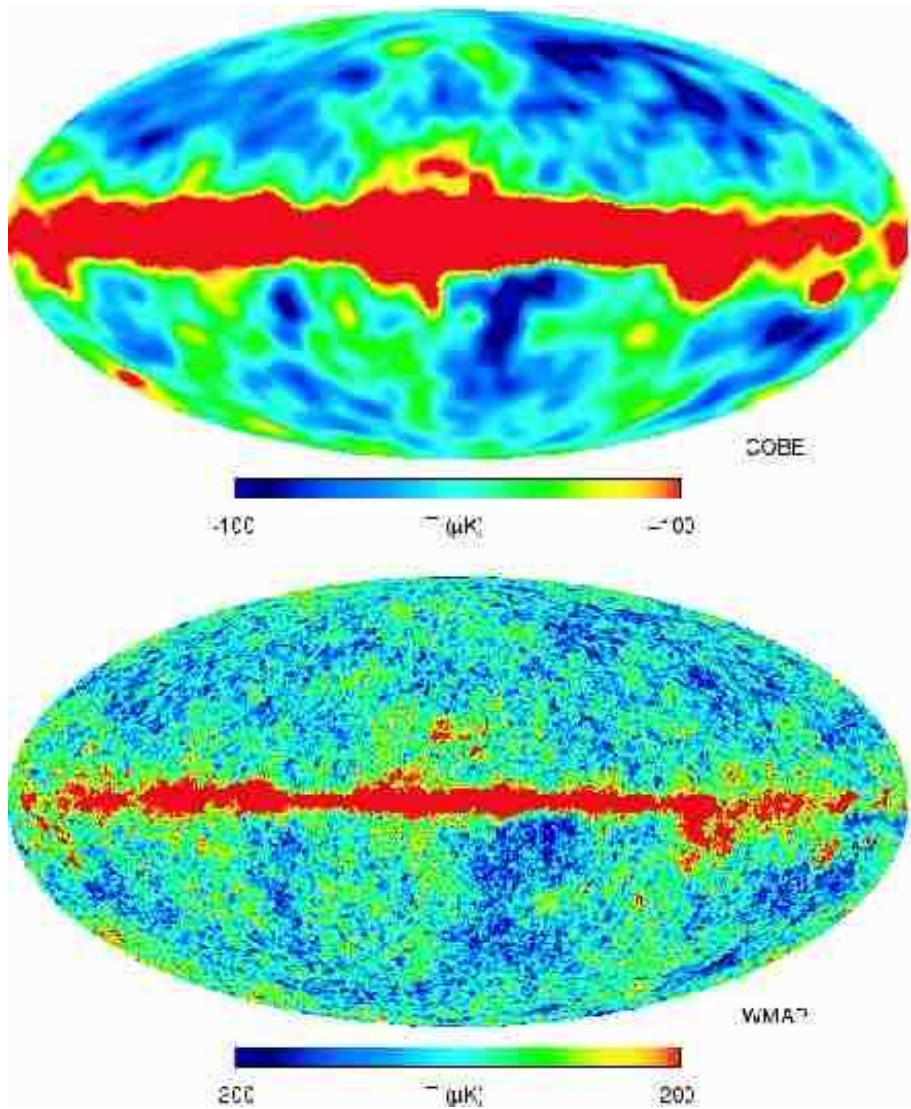}
\end{center}
\caption{{\it (Bottom)} The full-sky map of CMB anisotropies obtained by the
satellite WMAP in 2003 (see C.~L.~Bennett et al.,
Astrophys.J.Suppl. 148 (2003) 1). A higher-resolution image is
available at {\tt http://lambda.gsfc.nasa.gov/product/map/}.  The blue
(resp. red) spots correspond to colder (resp. warmer) regions.  The
central red stripe in the middle is the foreground contamination from
the Milky Way. {\it (Top)} The COBE map, shown again for comparision.
The resolution of WMAP is 30 times better than that of COBE, but one can 
easily
see that on large angular scales the two experiments
reveal the same structure.}
\end{figure}
\begin{figure}[!th]
\begin{center}
\epsfxsize=12cm
\epsfbox{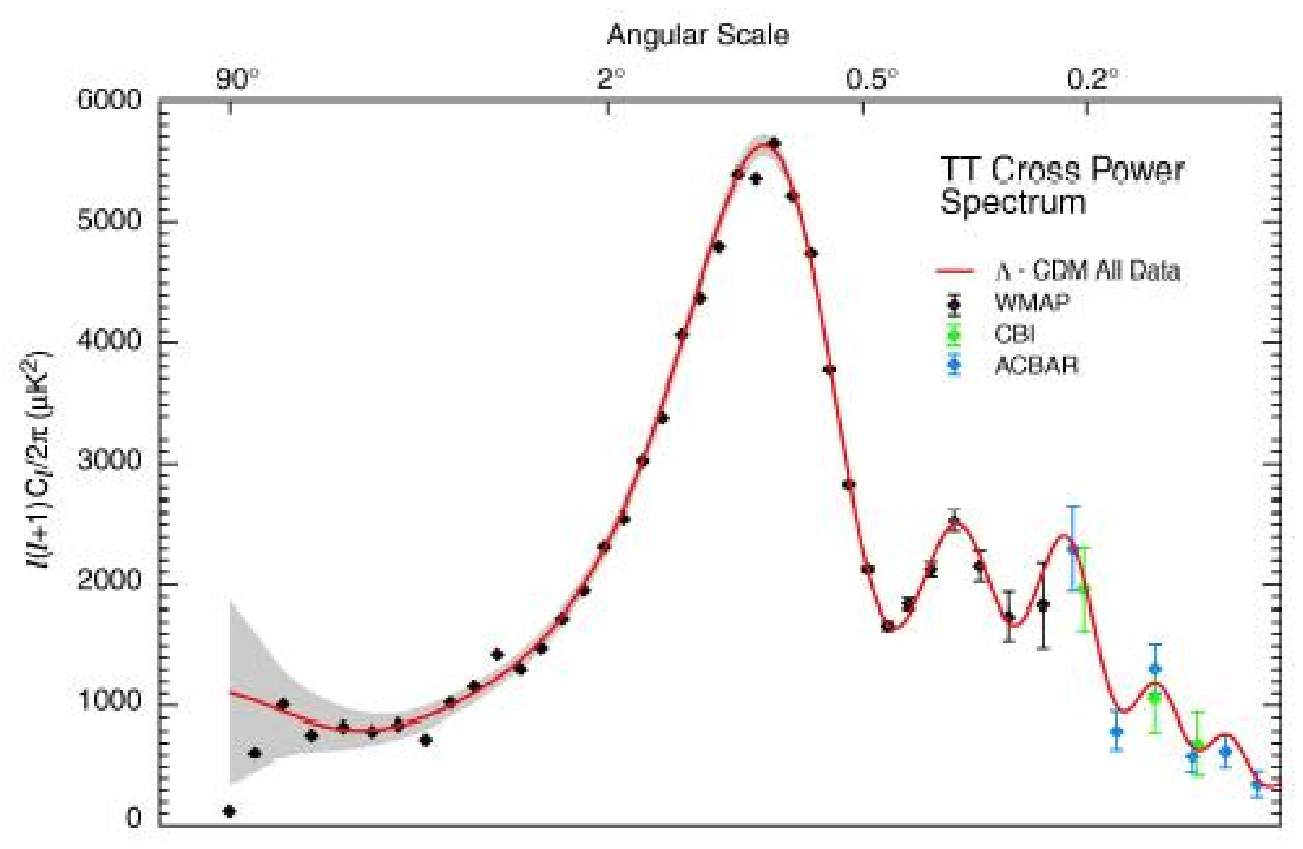}
\end{center}
\caption{ The spectrum of WMAP map is perfectly fitted by the
theoretical predictions, for a model called $\Lambda$CDM described in
section \ref{sec23}. The black dots show the WMAP measurements
(see  C.~L.~Bennett  et al., Astrophys.J.Suppl. 148 (2003) 1).  The
error bars are so small that they are difficult to distinguish from
the dots. The blue and green dots show some complementary measurements from
other experiments dedicated to smaller angular scales. The red curve is one of
the best
theoretical fits.
}
\end{figure}

There are still many CMB experiments going on. The satellite
WMAP goes on acquiring data in order to reach even better resolution.
In 2007, the European satellite
``Planck'' will be launched. Planck is expected to perform the best
possible measurement of the CMB temperature anisotropies, with such
great precision that most cosmological parameters should be measured
with only one per cent uncertainty.

\subsection{Matter perturbations}
\label{sec225}

We now focus on the evolution of the matter perturbations. Because we
believe that the Universe contains much more CDM than baryons, in
first approximation, we can forget the baryons, and do as if all the
matter was in the form of CDM, with absolutely no electromagnetic
interactions.  This is crucial, because in absence of any coupling with
the photons, the CDM fluid has a vanishing pressure: so, it tends to
concentrate in each well of gravitational potential -- instead of
oscillating like the baryons before decoupling.  
This is called gravitational collapse.

Of course, as all causal mechanisms, gravitational collapse can only affect
modes inside the Hubble radius. However, there is one more restriction.
During matter domination, when a potential well appears, CDM
concentrates in it, and deepens the gravitational potential even more.
So, both CDM and gravitational potential perturbations are quickly amplified.
On the other hand, during radiation domination, the gravitational potential 
follows
the perturbations of the dominant energy density: that of radiation.
So, there is no significant coupling between the CDM inhomogeneities
and the gravitational potential. As a consequence, the amplification
of CDM perturbations is efficient only {\it after} equality
and {\it inside} $R_H$. If we look at the diagram of figure 2.6,
this corresponds to the region on the right of the purple curve,
and below the yellow line.
So, even today, the Fourier spectrum of CDM perturbations
has a characteristic scale corresponding
to the Hubble radius at equality. The same type of numerical code
which compute the spectrum of CMB anisotropies can also compute
the linear matter spectrum with great precision, for any choice
of cosmological parameters.

But linear perturbation theory is valid only as long as the
perturbations are small, $\delta_{\rm CDM}(t,k) \ll 1$. After, it is
no longer interesting to use a Fourier decomposition, because the
evolution of each mode is not independent of the others: so,
everything becomes much more complicated, and we can make some
predictions only from numerical simulations of the gravitational
collapse in real space. These are called an N-body simulations.

It is crucial to know at what time a given mode enters into the
non-linear regime, $\delta_{\rm CDM} (t, k) \sim 1$. If there was
absolutely no growth of CDM perturbations inside the Hubble radius
during radiation domination, this time would be the same for all small 
wavelengths, because they would grow at the same 
rate, starting from the same value.
But because there is a very slow amplification of
perturbations during this epoch, the most amplified wavelength are
the smallest ones -- since they entered earlier into the Hubble radius.
So, the smallest wavelengths are the first ones to
become non-linear. As a consequence, the smallest structures in the Universe 
are the older ones: 
this is called hierarchical structure formation.

\subsection{Hierarchical structure formation}
\label{sec226}

Let's go back to equality, when no efficient amplification of the matter
perturbations has started: the matter density in the Universe is still
a continuous function, with small inhomogeneities of order $10^{-5}$.
Then, when the smallest wavelengths  become non-linear 
($\delta_{\rm CDM}(t,k) \sim 1$), some small compact clumps start to form.
But at that time, if we smooth the density distribution
with a smoothing radius corresponding to
the largest non-linear scale, we find again a continuous distribution
(remember the analogy in figure 1.4),
with a Fourier spectrum obeying to the predictions of the
linear theory of perturbations.

As time goes on, some larger and larger scales become non-linear and
start to collapse. So, some small overdense regions (protogalaxies)
are expected to form first; then, they merge into bigger and bigger
galaxies, and later clusters of galaxies will also appear. 
Finally, in the present Universe, the scale of non-linearity
is expected to be around 20 or 40 Mpc. This is one order of magnitude
bigger than clusters of galaxies: so, even the clusters are not
smoothly distributed. In fact, they tend to organize themselves along
huge filaments, separated by voids of size 20 to 40 Mpc. The structure
of the Universe on these huge scales looks very much like a spounge...
But by smoothing the present distribution of matter over
scales larger than 40~Mpc, one finds again a continuous distribution,
with small inhomogeneities described by the linear Fourier spectrum 
of cosmological perturbation theory.

\subsection{Observing the matter spectrum}
\label{sec227}

\begin{figure}[!th]
\begin{center}
\epsfxsize=12cm
\epsfbox{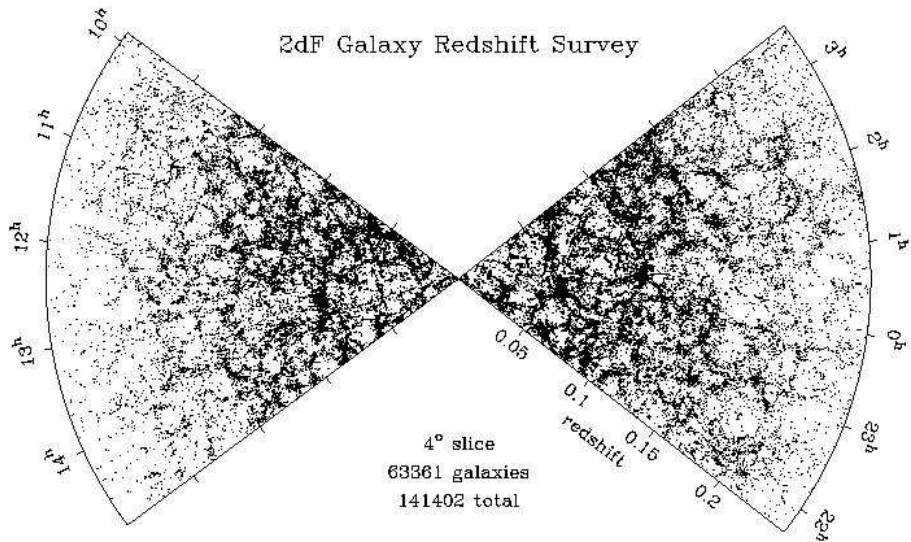}
\end{center}
\caption{The distribution of galaxies in two thin slices of the
neighboring Universe, obtained by the 2dF Galaxy Redshift Survey
(see J.~A.~Peacock et al., Nature 410 (2001) 169-173).  The
radial coordinate is the redshift, or the distance between the object
and us.}
\end{figure}
\begin{figure}[!th]
\begin{center}
\epsfxsize=12cm
\epsfbox{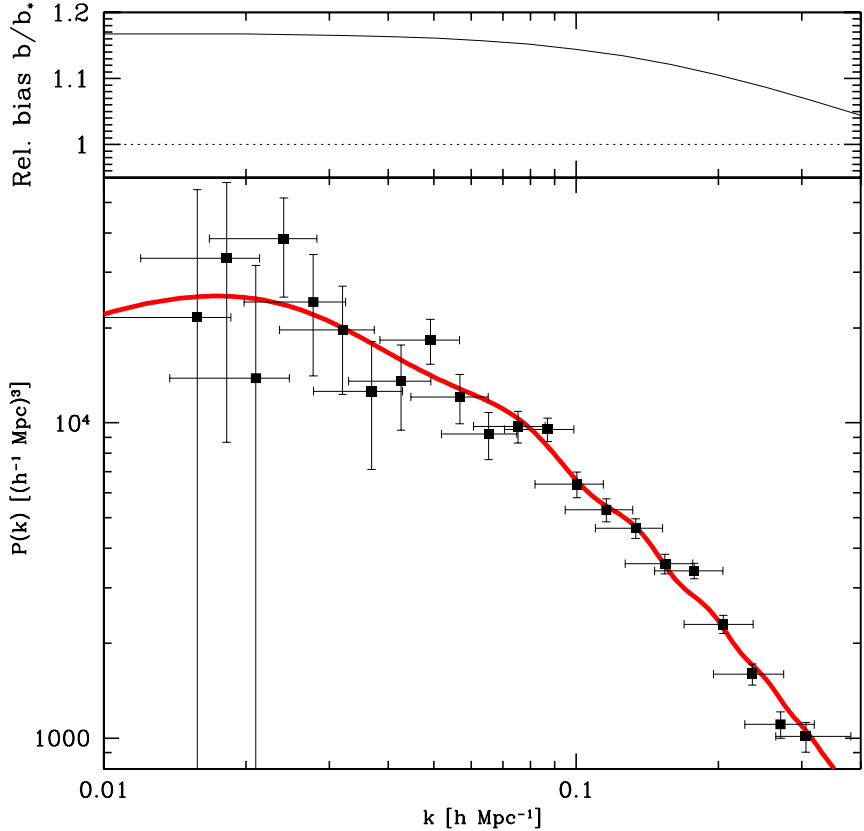}
\end{center}
\caption{
The Fourier spectrum of matter perturbations, reconstructed
from a large galaxy redshift survey (here, the Sloan Digital Sky Survey,
see M.~Tegmark et al., Astrophys.J.606 (2004) 702-740). 
In the lower plot,
the data (black squares) are compared with a theoretical spectrum
(red curve), obtained
by integrating numerically over a system of differential equations,
accounting for the evolution of linear perturbations.
In the calculation, the cosmological parameters were chosen in
order to obtain a maximum likelihood fit to the data.
Comparing the data with the linear spectrum is 
meaningful only for the largest wavelengths, $k < 0.1 h$~Mpc$^{-1}$,
which have not been affected by the non--linear evolution.
For smaller scales, the agreement between the data and
the theoretical prediction is only a coincidence. 
}
\end{figure}
So, if we have a large three--dimensional map of the galaxy
distribution in the Universe, we can smooth it on very large scales,
compare with the linear spectrum predicted by the theory, and measure
the cosmological parameters on which this spectrum depends.  Of
course, what we observe in the sky in a galaxy survey
is only luminous baryonic matter,
not CDM.  But astrophysicists have very good reasons to believe that
on large scales, the distribution of luminous matter ``traces'' the
distribution of mass. This means that the inhomogeneities in luminous
matter would be proportional to the total density inhomogeneities --
i.e., in good approximation, to the CDM inhomogeneities.  The
unknown coefficient of proportionality is called the bias $b$:
\begin{equation}
\delta_{\rm luminous}(t,k) \simeq b~ \delta_{\rm CDM}(t,k).
\end{equation}

Over the past decades, astronomers could build some very large
three-dimensional maps of the galaxy distribution -- even larger than
the scale of non-linearity.  On figure 2.14, we can see two thin
``slices'' of the surrounding Universe.  Inside these slices, each
point corresponds to a galaxy seen by the 2-degree Field Galaxy
Redshift Survey (2dF GRS). The radial coordinate is the redshift, or
the distance between the object and us. Even at redshift $z<0.1$, we
can see some darker regions, corresponding to the voids - the largest
non-linear structures in the Universe.  At $z>0.1$, the average
density seems to decrease smoothly, but this is simply because the
survey is limited in sensitivity -- at small distance it sees all the
galaxies, while at large distance it sees only the brightest ones.  By
taking this selection effect into account, and by smoothing over
distances of order of 50~Mpc, one recovers a very homogeneous
distribution of matter.

On figure 2.15, we see the Fourier spectrum reconstructed from another
redshift survey, the Sloan Digital Sky Survey (SDSS).
Each experimental point is obtained by smoothing over smaller 
wavelengths. The data points with $k < 0.1 h$~Mpc$^{-1}$ can be directly
compared with the linear spectrum predicted by the linear theory
of cosmological perturbations.
The red line shows one of the best-fit theoretical curves. From
such a comparison, one can extract some information on the value of the
cosmological parameters. This is similar to the parameter extraction from
CMB anisotropies, but it provides some independent constraints.

On the extreme left of figure 2.15, we see that the theoretical curve
reaches a maximum (around $k \sim  0.01h$~Mpc$^{-1}$, i.e.,
$\lambda \sim 400$~Mpc). This corresponds to the scale which entered
into the Hubble radius at the time of equality between radiation and matter.
Future redshift surveys
are expected to give better data around this scale. Observing the
turnover of the matter spectrum with good precision will
improve the measurement of the cosmological parameters.

In addition, astronomers are currently 
developing some new techniques for measuring
the large--scale power spectrum, based on gravitational lensing effects. These
methods measure directly the gravitational potential distribution
-- instead of luminous matter perturbations -- and are expected to produce
some very interesting results in the next decades.


\section{Measuring the cosmological parameters}
\label{sec23}

There has been some continuous progress in measuring the cosmological
parameters over the past three decades; however, the most spectacular
improvements have been made since 1998. Many observational strategies
can be employed for constraining the cosmological scenarios --
probably around twenty or thirty. Instead of giving an exhaustive
review, we will focus only on five important methods.

\subsection{Abundance of primordial elements}
\label{sec231}

In sections \ref{sec213} and \ref{sec215}, we have seen that the
theory of Nucleosynthesis can predict the abundance of light
elements formed in the early Universe, when the energy density was
of order $\rho \sim (1 \, {\rm MeV})^4$. This theory has reached today
a great level of precision: it is believed that each relevant nuclear 
reaction are taken carefully into account. In the standard model
of Nucleosynthesis, there are only two free parameters: the baryon
density $\rho_{\rm B}$, or equivalently, $\Omega_{\rm B} h^2$; and
the radiation density $\rho_{\rm R}$, or equivalently, $\Omega_{\rm R} h^2$.
Under the standard assumption that the radiation fluid is composed of
photons and of a number $N_{\nu}$ of neutrino families, alll of them
being in thermal equilibrium in the Early Universe,
we can
express $\Omega_{\rm R} h^2$ in terms of $N_{\nu}$.

For each
value of $\Omega_{\rm B} h^2$ and of $N_{\nu}$, the theory predicts a 
given abundance of
hydrogen, helium, lithium, etc. So, by observing the abundance of these
elements in the Universe -- for instance, using spectroscopy --
it is possible to constrain the baryon density. Current
observations lead to the prediction
\begin{equation}
\Omega_{\rm B} h^2 = 0.021 \pm 0.005.
\end{equation}
Moreover, they are perfectly compatible with $N_{\nu}=3$, 
which is known from accelerator physics to be the actual number of 
flavor neutrino
families.

\subsection{CMB anisotropies}
\label{sec232}

The precise dynamics of the acoustic oscillations before
decoupling depends on almost all of the cosmological parameters: 
so, the spectrum of CMB anisotropies can give a wide variety of 
constraints. This
type of experiment is considered as the most complete and precise
cosmological test. Future CMB experiments like Planck should
be able to measure each parameter with exquisite precision.
At the moment, some particular parameters or combinations of parameters
are already measured unambiguously, using the CMB anisotropy map of
WMAP
(and also of other experiments like ACBAR and CBI). One example
is the baryon density, which is related in an interesting way
(that we will not detail here) to the relative amplitude of the
acoustic peaks. The WMAP data gives
\begin{equation}
\Omega_{\rm B} h^2 = 0.023 \pm 0.002.
\end{equation}
This constraint is
in perfect agreement with that from Nucleosynthesis -- which is really 
striking, since these two measurements
are done with completely different techniques:
one uses nuclear physics, applied to the Universe when
it had a density of order (1~Mev)$^4$, while the other uses 
a general relativistic extension of fluid mechanics, 
applied to a much more recent era with density (0.1 eV)$^4$.
The perfect overlap between the two constraints indicates
that our knowledge of the Universe is 
impressively good, at least during the epoch 
between $t \sim 1$~s and
$t \sim 100~000$~yr.

We have seen that the position of the acoustic peak
is given by the Hubble radius at decoupling, which value is
predicted by the theory. So, using the angular 
diameter--redshift relation (see section \ref{sec135}),
it is possible to extract some information both on the spatial
curvature and on the evolution of the scale factor $a(t)$.
A detailed study shows that the leading effect is that
of spatial curvature. So, the position of the acoustic peak
gives a measurement of $\Omega_0 = 1 - \Omega_k$. The result is 
\begin{equation}
\Omega_0 = 1.03 \pm 0.05.
\end{equation}
So, according to current CMB experiments, the Universe
seems to be spatially flat. This result is extremely useful,
but it doesn't say how the present energy density
is distributed between matter and
a possible cosmological constant\footnote{As far as radiation is concerned,
we know that today, it is negligible with respect to matter, so
that $\Omega_0 \simeq \Omega_{\rm M} + \Omega_{\Lambda}
= \Omega_{\rm B} + \Omega_{\rm CDM} + \Omega_{\Lambda}$.}.

Current CMB experiments also give other interesting constraints (for
instance, on the primordial spectrum of perturbations) that we will not
explain here.

\subsection{Age of the Universe}
\label{sec233}

Observations can set a reliable lower bound on the age of the Universe:
if an object is seen at a given redshift, corresponding to a given
age, the Universe has to be older than this object.
During the 1990's, it has been established that the most distant observable
quasars are at least 11~Gyr old. On the other hand, a spatially flat
matter--dominated Universe would be only
9~Gyr old (see section \ref{sec216}). So, distant quasars seem
to indicate that
the Universe is either open or $\Lambda$--dominated today.

\subsection{Luminosity of Supernovae}
\label{sec234}

\begin{figure}[!th]
\begin{center}
\epsfxsize=12cm
\epsfbox{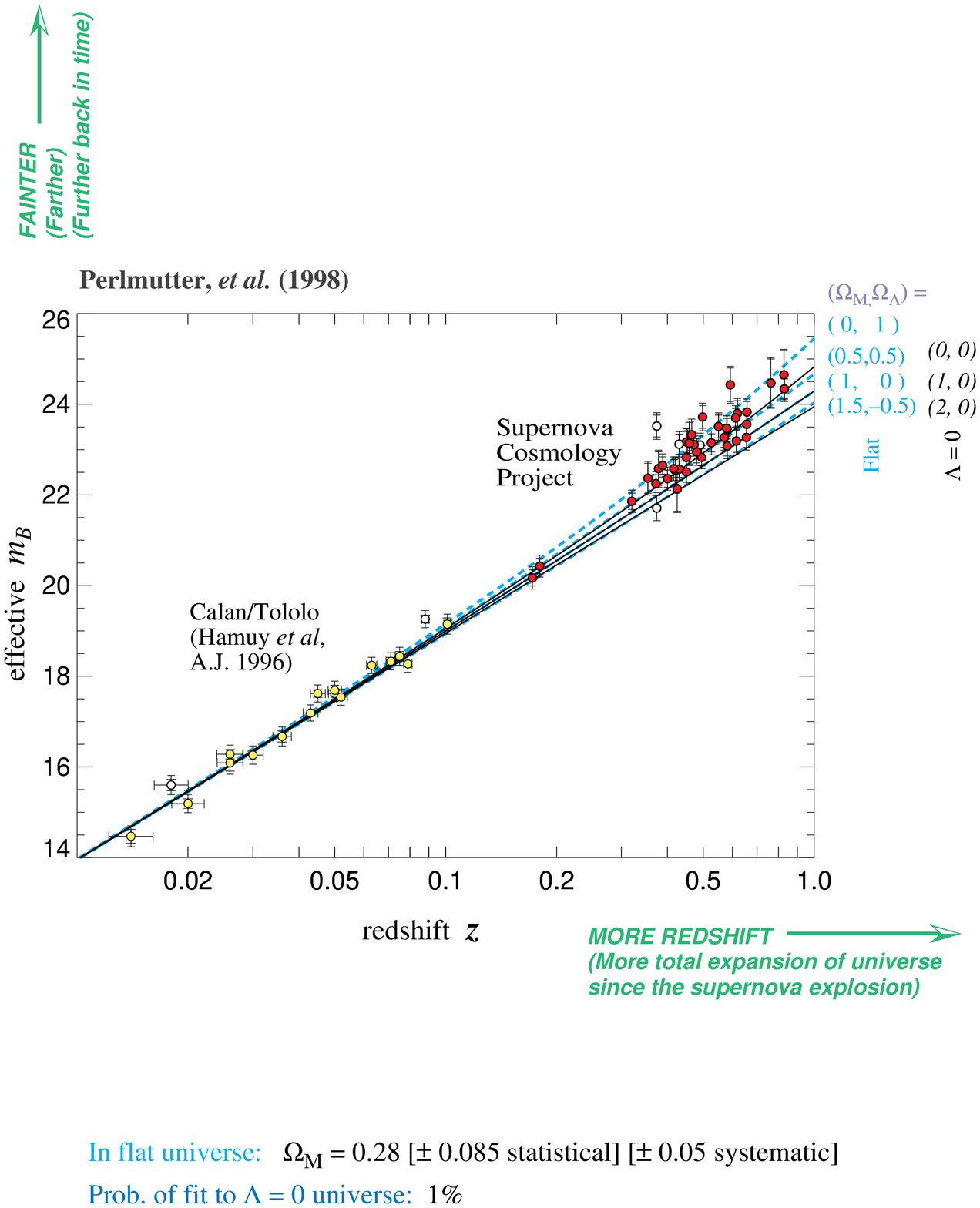}
\end{center}
\caption{The results published by the ``Supernovae Cosmology Project''
in 1998 (see Perlmutter et al., Astrophys.J. 517 (1999) 565-586).  
The various curves represent the effective
magnitude--redshift relation, computed for various choices of
$\Omega_{\rm M}$ and $\Omega_{\Lambda}$. This plot is equivalent to a
luminosity distance -- redshift relation (effective magnitude and
luminosity distance can be related in a straightforward way:
$m_B \propto (\log[d_L] + {\rm cst})$). 
The solid black curves
account for three examples of a closed/flat/open Universe with no
cosmological constant. The dashed blue curves correspond to three
spatially flat universes with different values of $\Omega_{\Lambda}$.
For a given value of $H_0$, all the curves are asymptotically equal at
short distance, probing only the Hubble law. The yellow points are
short--distance SNIa's: we can check that they are approximately
aligned.  The red points, at redshifts between 0.2 and 0.9, show that
distant supernovae are too faint to be compatible with a flat
matter--dominated Universe ($\Omega_{\rm M}, \Omega_{\Lambda}$) =(1,0).
}
\end{figure}
\begin{figure}[!bt]
\begin{center}
\epsfxsize=12cm
\epsfbox{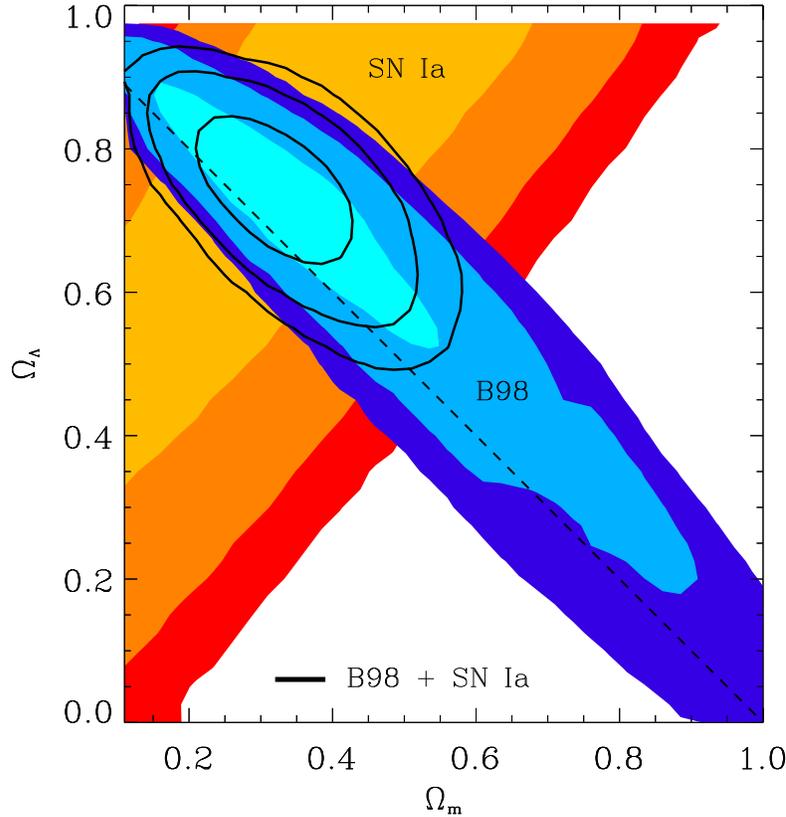}
\end{center}
\caption{
Combined constraints on ($\Omega_{\rm M}, \Omega_{\Lambda}$) from CMB
and Supernovae experiments (in 2001, 
see S.~Masi et al., Prog.Part.Nucl.Phys. 48 (2002) 243-261). 
The simple case of a spatially flat
Universe with no cosmological constant corresponds to the lower right
corner of the plot. In blue, the CMB allowed region favors
a spatially flat Universe with $\Omega_{\rm M} + \Omega_{\Lambda} \simeq 1$.
In orange, the Supernovae constraint is orthogonal to the CMB one.
By combining the two results, one obtains the favored region limited
by the black ellipses, centered near $(\Omega_{\rm M}, \Omega_{\Lambda})
= (0.3, 0.7)$.
}
\end{figure}

The evidence for a non--flat Universe and/or a non--zero cosmological
constant has increased considerably in 1998, when two independent
groups studied the apparent luminosity of distant type Ia supernovae
(SNIa). For this type of supernovae, astronomers believe that there is
a simple relation between the absolute magnitude and the luminosity
decay rate. In other words, by studying the rise and fall of the
luminosity curve during a few weeks, one can deduce the absolute
magnitude of a given SNIa. Therefore, it can be used in the same way
as cepheid, as a probe of the luminosity distance -- redshift
relation.  In addition, Supernovae are much brighter that cepheids,
and can be observed at much larger distances (until redshifts of order
one or two). While observable cepheids only probe short distances, where the
luminosity distance -- redshift relation only gives the Hubble law
(the proportionality between distance and redshift), the most distant
observable SNIa's are in the region where general relativity
corrections are important: so, they can provide a measurement of the
scale factor evolution (see section \ref{sec135}).

On figure 2.16, the various curves represent the effective
magnitude--redshift relation, computed for various choices of
$\Omega_{\rm M}$ and $\Omega_{\Lambda}$. The effective magnitude $m_B$
plotted here
is essentially equivalent to the luminosity distance $d_L$, since
it is proportional to $\log[d_L]$ plus a constant.  For
a given value of $H_0$, all the curves are asymptotically equal at
short distance. Significant differences show up only at redshifts
$z>0.2$. Each red data point corresponds to a single supernovae, from
the 1998 results of the ``Supernovae Cosmology Project''.  Even if it
is not very clear visually from the figure, a detailed statistical
analysis reveals that a flat matter--dominated Universe (with
$\Omega_{\rm M}=1$, $\Omega_{\Lambda}=0$) is excluded. This result has
been confirmed by more recent data at larger redshift. According to
statistical analysis, the allowed region in the ($\Omega_{\rm M},
\Omega_{\Lambda}$) parameter space is
\begin{equation}
-1 < \Omega_{\rm M} - \Omega_{\Lambda} < 0.
\end{equation}

When this observation is combined with the CMB results (figure 2.17), 
it appears that our Universe is spatially flat and has a non--zero 
cosmological constant -- a case corresponding to accelerated
expansion today! 
The favored values are ($\Omega_{\rm M}, \Omega_{\Lambda}$) = (0.3,0.7).

\subsection{Large Scale Structure}
\label{sec235}

We have seen that the Fourier spectrum of matter perturbations
on very large scales (figure 2.15) depends on the cosmological parameters.
Current galaxy redshift surveys provide
some constraints which are essentially independent from those of the
previous sections. It is remarkable that they converge toward the same
values as the previous techniques.

\vspace{0.5cm}

In conclusion, current experiments agree on the following matter budget
in the present Universe:
\begin{center}
\begin{tabular}{|c|}
\hline 
$
\Omega_0 = \Omega_{\rm B} + \Omega_{\rm CDM} + \Omega_{\Lambda} \simeq 1
$
\\
$
\Omega_{\Lambda} \simeq 0.73
$
\\
$
\Omega_{\rm B} \simeq 0.04
$
\\
$
\Omega_{\rm CDM} \simeq 0.23
$
\\
\hline
\end{tabular}
\end{center}
So, the Universe seems to be accelerating today, and 96\%
of its energy density is non--baryonic and of essentially unknown nature!
The measured value of the cosmological constant
shows that the Universe started to be $\Lambda$--dominated only
recently: at redshifts $z \geq 2$, the Universe
was still matter--dominated. We seem to be leaving precisely 
during the period of 
transition between matter--domination and $\Lambda$--domination!


\section{The Inflationary Universe}
\label{sec24}

\subsection{Problems with the Standard Cosmological Model}
\label{sec241}

According to the modern version of the standard cosmological model,
called the $\Lambda$CDM model,
the Universe is spatially flat, with three stages: radiation--domination
before $z \sim 10^4$,
matter--domination, and now, transition to $\Lambda$--domination.
This scenario is impressively successful in explaining a wide variety of
observations, as we could see in section \ref{sec23}.
However, it also raises a few problems.

\vspace{0.5cm}

{\bf The flatness problem.}
\\
The present contribution of the spatial curvature to the
expansion is given by $\Omega_k = \rho_0(t_0)/\rho_c(t_0) - 1$. 
But this contribution has evolved
with time.
From the Friedmann law, it is straightforward to see that
\begin{equation}
| \Omega_k(t) | = \left| \rho_0(t)/\rho_c(t) - 1 \right| 
= \frac{c^2 |k|}{a^2 H^2}
= \frac{c^2 |k|}{\dot{a}^2}.
\label{flatness}
\end{equation}
During the domination of
radiation (resp. matter), the factor $\dot{a}^{-2}$ 
grows like $t$ (resp. $t^{2/3}$).
Therefore, a spatially flat Universe is unstable: if initially,
$k$ is {\it exactly} zero,
$\Omega_k$ 
will remain zero; but if it is only {\it close} to zero, $\Omega_k$ will
increase with time! In order to have no significant curvature--domination
today ($| \Omega_k(t_0)| \leq 0.1$), we find that near the Planck time,
the curvature was incredibly small, roughly $| \Omega_k(t) | \leq 10^{-60}$...
The standard cosmological model provides no explanation for
obtaining such a small number in the early Universe: 
the fact that the Universe is spatially flat 
-- even approximately --
appears like the most unnatural possibility!
This is called
the ``flatness problem''.

\vspace{0.5cm}

{\bf The horizon problem.}  \\ 
We have seen that during radiation and
matter domination, the causal horizon is equal to the Hubble radius
(times a factor 2 or 4).  From the observation of CMB anisotropies, we know
that the Hubble radius at the time of decoupling is seen today under
an angle of one degree in the sky. So, on CMB maps, two points seen
under an angle bigger than one degree seem to be be causally
disconnected. In other words, the CMB maps seem to be composed of
approximately $10^3$ independent regions, which could not exchange any 
information in the past. But then, how can they be at
the same temperature -- up to $10^{-5}$ ? Given these theoretical arguments,
we would rather
expect each of these regions to have its own temperature. Then,
today we would observe inhomogeneities of order one.
Measuring approximately the same temperature $T_0=2.73$~K 
all over the sky appears like a
big mystery called the ``horizon problem''.

\vspace{0.5cm}

{\bf The origin of fluctuations.}
\\
The horizon problem can be reformulated in a slightly different way.
We have seen on figure 2.5 that during radiation and matter domination,
the Hubble radius (and the causal horizon) grow faster than the physical
wavelength of each perturbation.
So, all the wavelengths observable today on cosmological scales
were larger than the horizon in the early Universe. So far, we did not
propose any mechanism for the generation of primordial perturbations. 
But we have the following problem: if primordial perturbations are acausal, 
how can they be generated without violating causality (i.e., the fact
that no information can travel faster than the speed of light)?

\vspace{0.5cm}

{\bf The Cosmic Coincidence problem.}
\\
We have seen that the cosmological constant starts to dominate the
energy density of the Universe only in the very near past. Why are
we leaving just at the time of the transition? 
If $\Omega_{\Lambda}$ is of order one, then
$\rho_{\Lambda}^{1/4}$ is of order 10$^{-3}$~eV. Any vacuum energy
density resulting from a spontaneous symmetry breaking
in the standard model of particle physics 
should be at least of the order of the
electroweak scale, near 1~TeV. Eventually, some  
unknown
symmetry could force the vacuum energy to vanish exactly.
But why should it
be as small as 10$^{-3}$~eV? 

So, particle physics seems to be easily
compatible with a Universe that would be either completely matter-dominated or
completely $\Lambda$--dominated today. But the subtle balance
observed today between ordinary matter and the cosmological constant appears
as extremely unnatural and mysterious.

\subsection{An initial stage of inflation}
\label{sec242}

So, the standard cosmological model raises a series of problem -- the previous
list is not exhaustive. We will see that many of them can
be solved in an elegant way. 

Let us assume that in the early Universe,
before radiation domination, there has been a stage of accelerating expansion:
$\ddot{a}>0$.
So far, we don't say which mechanism could be responsible for
this period, called {\it inflation}.
If $\ddot{a}$ is positive, $\dot{a}$ increases, 
and according to equation (\ref{flatness}),
$|\Omega_k|$ decreases. 
The contribution of the spatial curvature to the 
expansion is driven to zero. This provides a possible solution to
the {\bf flatness problem}: during the initial inflationary stage,
$|\Omega_k|$ becomes extremely small. Then, it grows again, but today
it might still be very far from order one.

The simple relation that we found between the causal horizon and the 
Hubble radius is not true anymore during inflation. 
As a particular example of accelerated expansion, 
let us assume that $a(t)$ grows exponentially:
\begin{equation}
a(t) = e^{H_i t}.
\end{equation}
In that case, the Hubble radius is constant:
\begin{equation}
R_H = c \left( \frac{\dot{a}}{a} \right)^{-1} = \frac{c}{H_i},
\end{equation}
while the causal horizon grows exponentially:
\begin{equation}
d_H(t_1, t_2) = 2 a(t_2) \int_{t_1}^{t_2}  \frac{c~dt}{a(t)}
\longrightarrow  \frac{c}{H_i} e^{H(t_2-t_1)}
\qquad {\rm when} \qquad
t_2 \gg t_1.
\end{equation}
So, at the end of an inflation, the causal horizon is much
bigger than the Hubble radius. After, during radiation and matter
domination, they grow at the same rate, and 
the causal horizon {\it remains} larger.
So, an initial  stage of inflation can solve the {\bf horizon problem}: 
at the time 
of decoupling, if the causal horizon is at least on thousand times larger than 
the Hubble radius, it encompasses the whole observable Universe.
Then, all the points in the CMB map are in causal contact, 
and may have acquired the same temperature in the early Universe.

In the inflationary scenario, it is important not to make confusion
between the different causal horizons.  An horizon is always defined
with respect to an initial time.  We have seen that if the horizon is
defined with respect to the initial singularity, or to the beginning of
inflation, then it can be very large.  But the horizon defined with
respect to any time {\it after the end of inflation} is still close to
$R_H$. So, the causal distance for a physical process starting {\it
after} inflation is unchanged. In particular, the characteristic scale
that appears in the evolution of cosmological perturbations 
(section \ref{sec222}) is still
the Hubble radius. Our previous statement that the largest observable scales
are given by the Hubble radius evaluated today remains also true.
In inflationary cosmology, one must be very
careful because there are two different causal horizons, which are
relevant for different problems.

Let's come back finally to the {\bf origin of fluctuations}.
We can update figure 2.5 by adding a preliminary stage of inflation.
In general, the ratio of an arbitrary physical wavelength
over the Hubble radius is given by
\begin{equation}
\frac{\lambda(t)}{R_H(t)} = \frac{2 \pi a(t)}{k} \frac{\dot{a}(t)}{c~a(t)}
=  \frac{2 \pi \dot{a}(t)}{c~k}
\end{equation}
So, under the simple assumption that $\ddot{a}>0$, we see that this ratio 
increases during inflation: 
each wavelength grows faster than the Hubble radius.
We conclude that each Fourier mode observable today on cosmological 
scales started
inside the Hubble radius, crossed it during inflation, and re-entered later
(see figure 2.18).
So, all these wavelengths were
smaller than the causal horizon at some initial time.
This helps a lot in finding a realistic mechanism for the generation of
fluctuations, as we will see in the next section.
\begin{figure}[!bt]
\begin{center}
\epsfxsize=12cm
\epsfbox{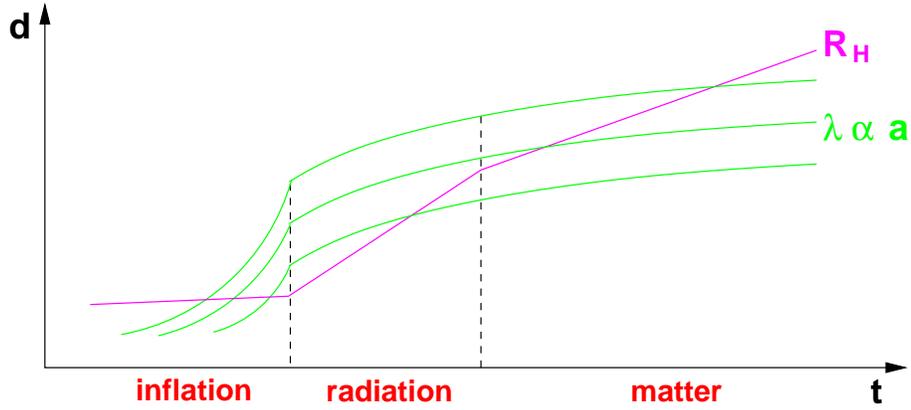}
\end{center}
\caption{
Comparison of the Hubble radius with the physical wavelength 
of a few cosmological perturbations. During the initial stage of accelerated
expansion (called inflation), the Hubble radius grows more slowly than each 
wavelength. So, cosmological perturbations originate from 
inside $R_H$. Then, the wavelength of each mode grows larger than 
the Hubble radius during inflation and 
re--enters during radiation or matter domination.
}
\end{figure}

\vspace{0.2cm}

All these arguments -- and also some other ones -- led a few cosmologists like
Guth, Starobinsky and Hawking
to introduce the theory of inflation in 1979.
Today, it is one of
the corner stones of cosmology. 

\subsection{Scalar field inflation}
\label{sec243}

So, the first three problems of section \ref{sec241} can
be solved under the assumption of a long enough stage of 
accelerated expansion in the early Universe. How can this
be implemented in practice?

First, by combining the Friedman equation (\ref{friedmann})
in a flat Universe with the conservation equation (\ref{conservation}),
it is easy to find that 
\begin{equation}
\ddot{a} > 0 \qquad \Rightarrow \qquad \rho + 3 p < 0.
\end{equation}
What type of matter corresponds to such an unusual relation
between density and pressure? A positive cosmological constant can do the
job:
\begin{equation}
p_{\Lambda} = - \rho_{\Lambda} \qquad \Rightarrow \qquad
\rho_{\Lambda} + 3 p_{\Lambda} = - 2 \rho_{\Lambda} <0.
\end{equation}
But since a cosmological constant is... constant, it cannot be responsible 
for an initial stage of inflation: otherwise this stage would go on
forever, and there would be no transition to radiation domination.

Let us consider instead the case of a scalar field 
(i.e., a field of spin zero, represented
by a simple function of time and space, and invariant under
Lorentz transformations). In general, for a homogeneous scalar field
$\varphi(t)$,
the density and pressure read
\begin{eqnarray}
\rho &=& \frac{1}{2} \dot{\varphi}^2 + V (\varphi), \\
p &=& \frac{1}{2} \dot{\varphi}^2 - V (\varphi),
\end{eqnarray}
where $V$ is the scalar potential. So, the pressure is equal to the kinetic 
energy {\it minus} the potential energy. We see that in the limit -- called
{\it slow--roll} -- where the kinetic energy is subdominant,
the pressure is close to $- \rho$, and the field behaves almost like a 
cosmological constant. However, the field is still slowly rolling 
toward the minimum of its potential -- which should have a {\it small}
slope, otherwise the kinetic energy would be leading.

So, a particular way to obtain a stage of accelerated expansion in the early 
Universe is to introduce a scalar field, with a flat enough potential. 
Scalar field inflation has been proposed in 1979 by Guth. During
the 80's, most important aspects of inflation were studied in details, by
Guth, Starobinsky, Hawking, Linde 
and other people. Finally, during the 90's,
many ideas and models were proposed in order to make contact
between inflation and particle physics. The purpose
of scalar field inflation is not only to provide a stage of accelerated
expansion in the early Universe, but also, a mechanism for the generation
of matter and radiation particles, and another mechanism for the
generation
of primordial cosmological perturbations. Let us summarize how it works
in a very sketchy way.

\vspace{0.2cm}

{\bf Slow-roll.} \\
\mbox{ }~~~~First, let us assume that just after the initial singularity, 
the energy density is dominated by a scalar field, with a potential
flat enough for slow--roll.
In any small region where the field is approximately homogeneous
and slowly--rolling, accelerated expansion takes place: this small region 
becomes
exponentially large, encompassing the totality of the present observable 
Universe. 
Inside this region, the causal horizon becomes much larger than the Hubble 
radius, and any initial spatial curvature is driven almost to zero -- so,
some of the main problems of the standard cosmological model are solved. After
some time, when the field approaches the minimum its potential, the 
slow-roll condition $\frac{1}{2} \dot{\varphi}^2 \leq V (\varphi)$ 
breaks down, and inflation ends: the expansion becomes 
decelerated again. 

\vspace{0.2cm}

{\bf Reheating.} 

At the end of inflation, the kinetic energy of the field is bigger 
than the potential energy; in general, the field is quickly oscillating
around the minimum of the potential. According to the laws of quantum 
field theory, the oscillating scalar field will decay into fermions and 
bosons. This could explain the origin of all
the particles filling our Universe.
The particles probably 
reach quickly a thermal equilibrium: this is why
this stage is called ``reheating''.

\vspace{0.2cm}

{\bf Generation of primordial perturbations.} 

Finally, the theory of scalar field inflation also explains the origin
of cosmological perturbations -- the ones leading to CMB anisotropies
and large scale structure formation. Using again quantum field theory
in curved space--time, it is possible to compute the amplitude of the
small quantum fluctuations of the scalar field.  The physical
wavelengths of these fluctuations grow quickly, like in figure
2.18. So, they are initially inside the Hubble radius, where quantum
mechanics tells us how to impose the initial normalization of each
mode.  The perturbations of the scalar field also generate some
perturbations of the space--time curvature -- in simple words, of the
gravitational potential.

When the wavelength of a mode of the gravitational potential
becomes bigger than the Hubble radius, the perturbation freezes out:
it is not affected by the decay of the scalar field during
reheating. But when the radiation and matter particles are formed
during reheating, they are sensitive to the gravitational potential,
and more particles accumulate in the potential wells. So, the
gravitational potential behaves like a mediator between the scalar
field perturbations during inflation and the radiation/matter
perturbations in the radiation/matter--dominated Universe. Through
this mechanism, all the primordial perturbations of photons, baryons and 
CDM originate from the quantum fluctuations of the
scalar field at the beginning of inflation.

Of course, the perturbations that we observe today are not of quantum
nature, and we don't need to employ wave functions and operators when
we compute the evolution CMB anisotropies.  This is consistent with
another nice property of inflation, which can be deduced rigorously
from the laws of quantum field theory, applied to the inflationary
space-time in accelerated expansion. During inflation, and after
crossing out the Hubble radius, each perturbation undergoes a
quantum--to--classical transition. In other words, quantum mechanics
is unavoidable in order to describe them initially, but later, they
become gradually equivalent to random classical fluctuations.

It is far beyond the level of these notes to compute the evolution of
primordial perturbations during inflation. However, we should stress
that it can be studied mathematically, in a very precise way. This
leads to generic predictions for the primordial perturbations, and in
particular, for their Fourier spectrum, which is related to the shape
of the scalar potential. The biggest success of inflation is that the
CMB anisotropies observed for instance by WMAP are in perfect
agreement with the predictions of inflation concerning the primordial
perturbations (in technical words, they are coherent, Gaussian,
adiabatic, and approximately scale--invariant: we will not explain
these concepts here). A priori, such a nice agreement between theory
and observations is far from obvious for a generic theory. It is
possible to build some mechanisms with an initial stage of accelerated
expansion, which solve the flatness and horizon problem, but fail in
explaining the spectrum of primordial perturbations. Current
observations of CMB anisotropies are the best evidence in favor of
scalar field inflation, and with future CMB experiments, it should be
possible to reconstruct the shape of the inflationary potential to
some extent; that will be really fantastic: by observing today a
picture of the Universe at the time of decoupling, we should infer some
details about what happened just after the initial singularity -- at
energies of order $\rho^{1/4} \sim 10^{16}$~GeV~!

\vspace{0.2cm}

Inflation is still a very active field of research, especially as far
as the connection with particle physics is concerned. First, the
scalar field invoked for inflation seems to be {\it ad hoc}: it is
introduced just for solving some cosmological problems, and does not
lead to new predictions concerning particle physics. However, there
are several proposals concerning the role of this field in particle
physics model. For instance, it could be a Higgs boson associated with
the GUT symmetry, or it could have to do with extra dimensions in
string theory. It is probably not impossible to test these assumptions, at
least in some indirect way (e.g., through the shape of the primordial
spectrum of perturbations). Second, the theory of reheating is still
being investigated in details, in a more an more realistic way from
the point of view of particle physics.

\subsection{Quintessence ?}
\label{sec244}

It is a pity that the theory of inflation, which is able to solve so many 
crucial issues in cosmology, doesn't say anything about the Cosmic 
Coincidence problem. Inflation predicts that 
$\Omega_{\rm M} + \Omega_{\Lambda}=1$,
in agreement with what is observed today, but it doesn't give any hint on the
splitting between matter and cosmological constant -- while the hint
from particle physics is either $\Omega_{\Lambda} = 0$ or 
$\Omega_{\Lambda} = 1$, at odds with observations.
As long as we explain the supernovae data with a cosmological constant,
it appears as an incredible coincidence that $\rho_{\Lambda}$ is of
the same order of magnitude as
$\rho_{\rm M}$ today, while
in the early Universe it was incredibly
smaller...

We have seen that during slow-roll inflation, the pressure of the scalar field
is negative, almost like for a cosmological constant. In the past five years, 
many people have proposed to explain the universe acceleration with a 
slow-rolling scalar field. So, there would be two stages of
scalar field inflation, one in the early Universe, and one today. A priori, 
they would be caused by two different scalar fields, called respectively the
inflaton and the quintessence field (although some models try to unify the two).
The big advantage with these models is that the energy density
responsible for the cosmic
acceleration today doesn't have to be constant in the past. 
In the early Universe,
it could very well be of the same order of magnitude as that of radiation,
and only slightly smaller.
During radiation domination, the density of quintessence and radiation
would decay at the same rate, and after radiation--matter 
equality, quintessence would enter into a stage of slow--roll. Then, its
energy density would remain almost constant, and it would start to 
dominate today, leading to the acceleration of the Universe.

Such classes of models are under investigation. However, they have
their own problems, and they don't solve the Cosmic Coincidence
problem in a fully convincing way. In fact, a wide variety of models
have been proposed in order to explain the supernovae data, but none
of them succeeds in explaining while the Universe starts to accelerate
only in the very near past. Together with the nature of dark matter,
the nature of the negative pressure term is the most puzzling feature
of modern cosmology -- and these two questions are far from being
small details, since altogether, $\Lambda$ and CDM account for 96\% of
the present energy density in the Universe! So, cosmology has done
some very impressive progresses in the past decades, but it seems that
the near future could be rich in other exciting discoveries!

\section*{Acknowledgments}

I would like to thank all the organizers and participants of the 2002,
2003 and 2004 schools for their stimulating interest, disturbing
questions and interesting comments, as well as some very kind
internauts for their help in improving this manuscript.
I apologize to the 2004 students for having had to miss the last courses,
and wish to thank Subir Sarkar for accepting to replace me on this occasion,
for their benefit!

\section*{Bibliography}

{[1]} {\it An Introduction to Modern Cosmology}, by Andrew Liddle, 
John Wiley \& Sons, Chichester, 2003. \\
{[2]} {\it Modern cosmology}, by Scott Dodelson,
New York, NY: Academic Press, 2003. \\
{[3]} {\it The Early Universe}, by Edward W. Kolb and Michael S. Turner,
Redwood City, CA: Addison-Wesley, 1990.

\end{document}